\documentclass{iau} 
\usepackage{graphicx}
\usepackage{natbib}
\usepackage{hyperref}

\pubyear{2022}
\volume{370}
\jname{Winds of Stars and Exoplanets}
\editors{A.~A. Vidotto, L.~Fossati \& J.~Vink, eds.}

\title[Observations of planetary winds and outflows] 
{Observations of planetary winds and outflows}

\author[L. A. Dos Santos]   
{Leonardo A. Dos Santos}

\affiliation{Space Telescope Science Institute, 3700 San Martin Drive, Baltimore, MD 21218, USA}

\begin{document}

\maketitle

\begin{abstract}
  We have recently hit the milestone of 5,000 exoplanets discovered. In stark contrast with the Solar System, most of the exoplanets we know to date orbit extremely close to their host stars, causing them to lose copious amounts of gas through atmospheric escape at some stage in their lives. In some planets, this process can be so dramatic that they shrink in timescales of a few million to billions of years, imprinting features in the demographics of transiting exoplanets. Depending on the transit geometry, ionizing conditions, and atmospheric properties, a planetary outflow can be observed using transmission spectroscopy in the ultraviolet, optical or near-infrared. In this review, we will discuss the main techniques to observe evaporating exoplanets and their results. To date, we have evidence that at least 28 exoplanets are currently losing their atmospheres, and the literature has reported at least 42 non-detections.
\keywords{(stars:) planetary systems, planets and satellites: general, techniques: spectroscopic}
\end{abstract}

\firstsection 
              
\section{Introduction}

The discovery of 51~Peg~b, a Jupiter-mass planet orbiting a Sun-like star with a period of only 4.3 days \citep{1995Natur.378..355M}, was initially received by the astronomical community with skepticism. But less than one year after this unexpected discovery, several other short-period gas giants were announced by competing teams \citep{1996Sci...273..429S}, leading us to come to terms with these so-called ``hot Jupiters" likely being a natural outcome of planet formation and evolution. One of the first questions that were posed during these early years of exoplanet science was whether hot Jupiters could survive the high mass loss rates driven by the extreme stellar irradiation at short periods \citep[][]{1996ApJ...459L..35G}. The current consensus is that hot Jupiters are massive enough to retain their atmospheres for billions of years, but the same cannot be said about other hot exoplanets \citep[e.g.,][]{2007A&A...461.1185L, 2007Natur.450..845K}.

Hydrodynamic atmospheric escape was originally formulated by \citet{1981Icar...48..150W} to explain the early evolution of the Earth and Venus. The idea behind this formulation stemmed from the insight of \citet{1972JAtS...29..214G}, who argued that, for planets with exospheres hotter than $\sim$10\,000~K, a selective escape of gases \citep[as in][]{1963GeoJ....7..490O} would be impossible. Instead, what follows is a bulk motion of gas in the upper atmosphere, or a so-called planetary outflow. More than two decades later, however, this process would be invoked to explain, at least partially, the observation of extended atmospheres in transiting exoplanets \citep[e.g.,][]{2003Natur.422..143V, 2010A&A...514A..72L} and later some demographic features in the exoplanet population \citep[e.g.,][]{2011ApJ...727L..44S, 2013ApJ...775..105O}. 

The main technique used to observe outflows in exoplanets is called transmission spectroscopy. This is the same method that yielded the first detection of sodium in an exoplanet \citep{2002ApJ...568..377C}, and remains one of the most prolific techniques to study atmospheres in extrasolar worlds. When a planet transits, part of the host star's light is filtered through the thin layer of gas at the limbs of the planet, imprinting wavelength-dependent signatures in the in-transit spectrum. This dependency emerges mainly due to a combination of the density, velocity, altitude and chemical composition of the absorbing material.

Tipically, the ratio of the area covered by the lower-atmosphere and the disk of a star is in the order of 10$^{-3}$ to 10$^{-4}$ in the optical and near-infrared \citep{2000ApJ...537..916S}. This level of precision requires strong spectroscopic features to be detectable \citep[e.g.,][]{2015A&A...577A..62W, 2016Natur.529...59S}. At higher altitudes, where the atmosphere is gravitationally unbound from the planet, the diffuse gas can extend to several planetary radii and produce deep in-transit absorption signatures \citep[e.g.,][]{2015Natur.522..459E}. It is precisely at high altitudes that we can observe signatures of planetary outflows.

This review has the following structure: In Sect. \ref{t_spec}, we will discuss the basic formulation that serves as the backbone of the transit spectroscopy technique; in Sect. \ref{escape_H}, we shall go over the main results of searches for escape of hydrogen using the Lyman-$\alpha$ and Balmer-series lines; in Sect. \ref{escape_metals}, we discuss the observations of exospheric metals, the smoking-gun signal of hydrodynamic escape in exoplanets; in Sect. \ref{escape_He}, we will discuss metastable He transmission spectroscopy, currently the most productive technique to observe atmospheric escape; finally, in Sect. \ref{future}, we draw some of the main conclusions stemming from these observations and propose some new perspectives for future research in this sub-field of exoplanet science.

\section{The basics of transit spectroscopy}\label{t_spec}

In this manuscript, we shall adopt that the transmission spectrum $\phi$ of an exoplanet in function of transit phase $\theta$ and wavelength $\lambda$ is given by:

\begin{equation}
  \phi(\theta, \lambda) = 1 - \frac{f_\mathrm{in}(\theta,\lambda)}{F_\mathrm{out}(\lambda)} \mathrm{.}
\end{equation}where $F_\mathrm{out}$ is the out-of-transit spectrum of the host star and $f_\mathrm{in}$ is the observed in-transit spectrum\footnote{\footnotesize{We use lower-case $f$ to denote the dependence of the in-transit spectra to the orbital phase $\theta$. For the out-of-transit flux, we adopt upper-case $F$ to indicate that it does not depend on the planetary phase.}}. It is also convenient to define the transmission spectrum in the rest frame of the planet by Doppler shifting the spectra according to:

\begin{equation}
  \lambda_\mathrm{p}(\theta) = \lambda \left(\frac{c}{\Delta v(\theta)} + 1 \right) \mathrm{,}
\end{equation}where $\Delta v$ is the difference between radial velocity of the planet at a particular phase $\theta$ and the reference velocity of the observer, and $\lambda_\mathrm{p}$ will be the resulting wavelength in the rest frame of the planet. Finally, we can define the transmission spectrum $\Phi$ independent from the planetary phase by taking the mean of $\phi(\theta, \lambda_\mathrm{p})$ over the range of $\theta$ observed in transit:

\begin{equation}
  \Phi (\lambda_\mathrm{p}) = \frac{1}{\Delta \theta}\int \phi(\theta, \lambda_\mathrm{p}) d\theta \mathrm{.}
\end{equation}At low spectral resolution we cannot resolve the variation of the in-transit absorption with respect to the planetary Doppler velocity, and the transmission spectrum can be simplified to:

\begin{equation}
  \Phi(\lambda) = 1 - \frac{F_\mathrm{in}(\lambda)}{F_\mathrm{out}(\lambda)} \mathrm{.}
\end{equation}

In the formulation described above, we averaged the in-transit signature over the phase space and study the signature in function of wavelength. As we will see in Sections \ref{escape_H} and \ref{escape_metals}, light curves are also routinely used to study transit spectra and search for in-transit excess absorption that could indicate the presence of an atmosphere. In this method, we instead average signals over the wavelength space, and analyze its dependence in fuction of transit phase. As a recommendation for the reader, a more detailed treatise on transit spectroscopy can be found in \citet{2021exbi.book....7D}.

\section{Escape of H: Lyman-$\alpha$ and Balmer-series spectroscopy}\label{escape_H}

Classically, observations of atmospheric escape in exoplanets have been performed in ultraviolet (UV), which probes escape of hydrogen (H) and metallic species (see Sect. \ref{escape_metals}). The spectral feature of strongest interest is the Lyman-$\alpha$ (hereafter Ly$\alpha$) line at 1215.67~\AA, which traces atomic H. The \emph{Hubble Space Telescope} (\emph{HST}) is currently the only instrument capable of observing the Ly$\alpha$, and it is possibly going to remain in this position until the launch of the next flagship NASA space telescope \citep{NAP26141}. 

Since the interstellar medium (ISM) is rich in neutral H, the stellar Ly$\alpha$ line is partially or completely absorbed when observed from the Solar System. For stars with low radial velocities, the ISM absorption takes place near the core of the line; those with large radial velocities in relation to the Solar System and the ISM manage to dodge the absorption, and their Ly$\alpha$ cores are observable; see, e.g., the cases of Kepler-444 \citep{Bourrier17d} and Barnard's Star \citep{2020AJ....160..237F}. Save a few exceptions, it is likely that the Ly$\alpha$ line is completely absorbed by the ISM for F, G and K-type stars beyond 60~pc; for M dwarfs, this limiting distance is much shorter.

Some other H features can be observed at optical wavelengths, such as the Balmer series (which include H$\alpha$, H$\beta$, H$\gamma$), and provide another window to observe H escape in exoplanets. Observing in the optical has its advantages: there is no strict need to use a space telescope, ISM absorption is not a limiting issue, and it can be performed at high resolution. The disadvantages are that only highly-irradiated hot Jupiters display an in-transit absorption in the Balmer series, and no detection has so far been obtained for smaller or less irradiated planets (see Sect. \ref{H_nondet}). 

\subsection{Hot Jupiters}

The first exoplanet to have a definitive detection of escaping H was the hot Jupiter HD~209458~b, as originally reported by \citet{2003Natur.422..143V} and later confirmed by \citet{2008A&A...483..933E}. Using \emph{HST} and the Space Telescope Imaging Spectrograph (STIS), \citet{2003Natur.422..143V} detected a flux decrease of $15\% \pm 4\%$ in the blue wing of the Ly$\alpha$ line of the host star during the transit of the planet; since the transit depth at optical wavelengths is only $\sim1.5\%$, the authors argued that the excess absorption seen in Ly$\alpha$ is due to a large cloud of H surrounding HD~209458~b, which in turn is fed by atmospheric escape. 

One particular point of contention in the literature related to Ly$\alpha$ detections is regarding the Doppler velocities at which the signatures are measured; in the case of HD~209458~b, the in-transit planetary absorption takes place at velocities as high as $-130$~km\,s$^{-1}$ in the stellar rest frame, indicating that the detected escaping material is accelerated away from the star. One-dimensional hydrodynamic escape models are unable to explain such high velocities \citep{2009ApJ...693...23M}, thus requiring other processes to explain them. The exact mechanism behind this effect has been the subject of an intense debate in the literature, and the most discussed contenders are radiation pressure and charge exchange in the interface between the stellar and planetary winds \citep[e.g.,][]{2008Natur.451..970H, 2008Natur.456E...1L, 2013A&A...557A.124B, 2017MNRAS.470.4026V, 2018ApJ...860..175W, 2020MNRAS.493.1292D}.

Using \emph{HST}, but this time with the 1-st order CCD/G430M setup at optical wavelengths, \citet{2007Natur.445..511B} reported on an excess absorption of $0.03\% \pm 0.006\%$ during the transit of HD~209458~b. The authors argue that this feature is caused by a large population of hot H atoms in the planet's upper atmosphere, which absorb the stellar light in the Balmer jump and continuum.

Another early discovery of evaporation was that of the extensively studied hot Jupiter HD~189733~b \citep{2010A&A...514A..72L}, for which the authors detect an in-transit Ly$\alpha$ absorption of $14.4\% \pm 3.6\%$. Similar to HD~209458~b, this absorption takes place at highly blueshifted Doppler velocities. The main point of discussion for this planet is that there is strong evidence that its escape signals and its high-energy environment are variable \citep{2012A&A...543L...4L, 2013A&A...551A..63B, 2020MNRAS.493..559B, 2022A&A...660A..75P}; as we shall see in the next sections, this variability has not only been observed in Ly$\alpha$, but other wavelengths as well. 

The search for escape of atomic H is complicated by the fact that they rely predominantly on \emph{HST}, which is oversubscribed. However, excited H has been detected in the archetypal hot Jupiters HD~209458~b and HD~189733~b \citep{2012ApJ...751...86J}, and in the ultra-hot Jupiter KELT-9~b \citep{2018NatAs...2..714Y, 2019AJ....157...69C, 2020A&A...638A..87W, 2022A&A...666L...1S} using the Balmer series H lines. Similar to the Ly$\alpha$ observations, HD~189733~b also displays signals of variability in the H$\alpha$ line \citep{2017AJ....153..217C}. Other hot Jupiters with reported H$\alpha$ detections are WASP-12~b \citep{2018AJ....156..154J}, KELT-20~b \citep{2018A&A...616A.151C}, WASP-52~b \citep{2020A&A...635A.171C}, WASP-33~b \citep{2021A&A...645A..22Y}, and WASP-121~b \citep{2021ApJ...907L..47Y}. 

One of the main differences between the Ly$\alpha$ and H$\alpha$ detections is that the latter tend to display excess in-transit absorption in the order of $1\%$, which is shallower than the former.  Another key difference is that ISM absorption is not a limitation for these observations, and we have access to the core of the absorption. In fact, observations at high spectral resolution show that the excess in-transit signals in the Balmer series do not show a net blueshift, and are thus confined to relatively low Doppler velocities when compared to Ly$\alpha$. From the modeling perspective, these low-velocity signatures are advantageous because they do not require expensive three-dimensional simulations. This, in turn, means that we can use simplified formulations to extract mass loss rates for the observed exoplanet, such as the Parker-wind \citep{1958ApJ...128..664P} approximation, as seen in \citet{2020A&A...638A..87W} and \citet{2021A&A...645A..22Y}.

\subsection{Neptunes and sub-Neptunes}

\begin{figure}[h]
\begin{center}
\includegraphics[width=0.7\textwidth]{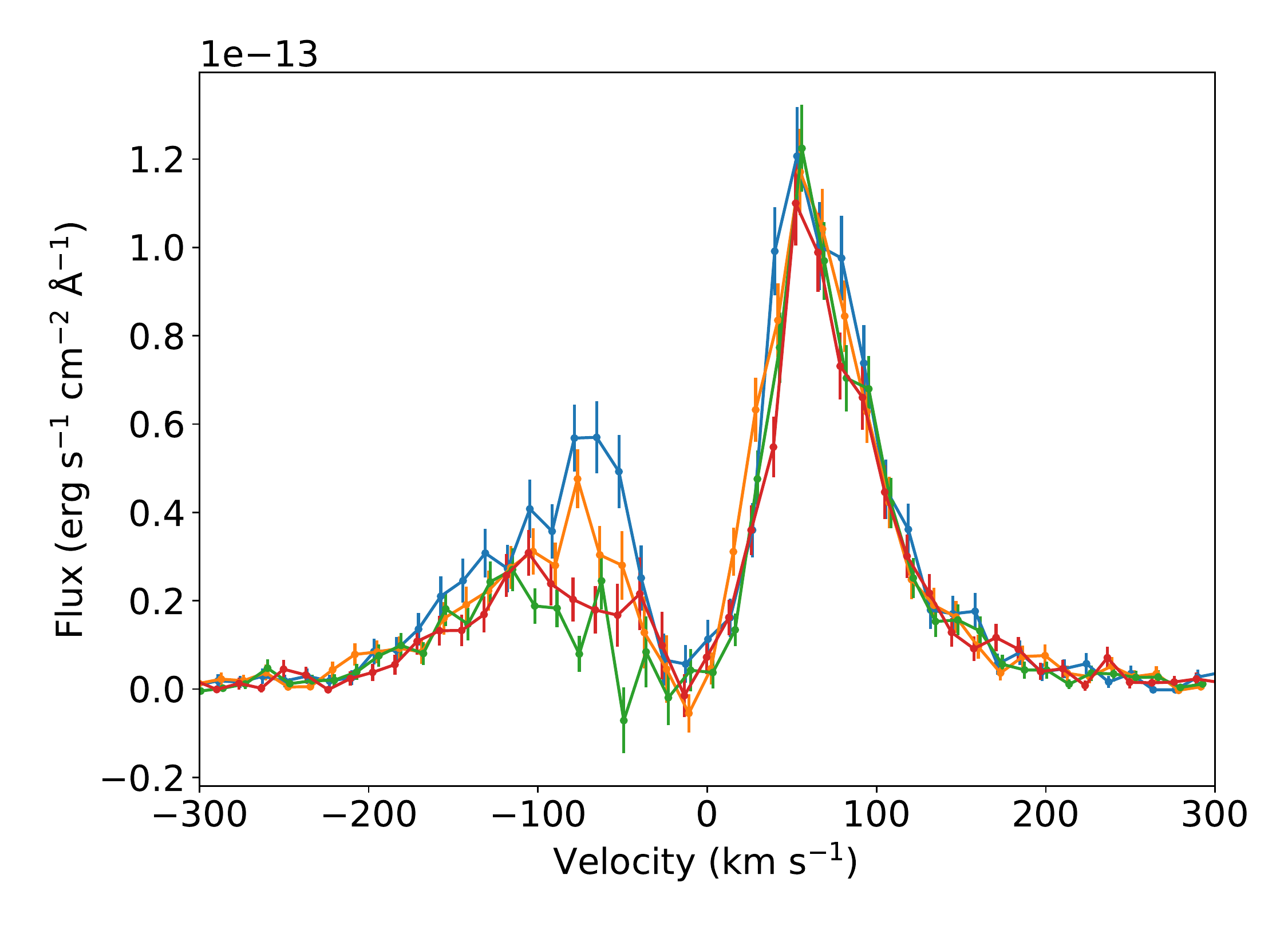} 
\caption{Lyman-$\alpha$ flux time-series during the transit of Gl~436~b, where blue and orange correspond to the spectra before the transit, green during the transit, and red after the transit \citep{2015Natur.522..459E}. The deep absorption in the blue wing, between Doppler velocities [-120, -50]~km\,s$^{-1}$, is explained by a large cloud of H around Gl~436~b fed by atmospheric escape.} 
\label{fig:gj436b_lya}
\end{center}
\end{figure}

Although the first observations of atmospheric escape in exoplanets were obtained for hot Jupiters, \citet{2011A&A...529A..80E} predicted that evaporating Neptune-sized worlds could not only be observed as well, but would show excess in-transit absorption just as deep as their larger counterparts. What they did not predict is that this signal could, in fact, be even larger than that. Upon observing the warm Neptune Gl~436~b (also known as GJ~436~b) with \emph{HST}/STIS, \citet{2015Natur.522..459E} found that the Ly$\alpha$ blue wing of the host star is obscured by a factor of $56.3\% \pm 3.5\%$ when the planet transits \citep[see Fig. \ref{fig:gj436b_lya}; see also][]{2014ApJ...786..132K}. Such a signal can only be explained by the presence of a large cloud of atomic H around the planet, fed by an atmospheric escape rate in the order of $10^9$~g\,s$^{-1}$ and accelerated away from the star. Further observations would later show that Ly$\alpha$ transit of Gl~436~b is not only deep, but also extremely asymmetric and long \citep{2017A&A...605L...7L}, stable across several years and observable with \emph{HST}/COS \citep{DosSantos19}. Since then, Gl~436~b has become the archetypal evaporating Neptune, and its observations have been extensively used to test modeling frameworks for atmospheric escape \citep[e.g.,][]{2015A&A...582A..65B, 2016A&A...591A.121B, 2019A&A...623A.131K, 2019ApJ...885...67K, 2021MNRAS.501.4383V, 2021A&A...647A..40A, 2021MNRAS.500.3382C}. Perhaps another warm Neptune that has become almost as iconic as Gl~436~b in the last few years is HAT-P-11~b, which displays signatures of atmospheric escape not only in Ly$\alpha$ \citep{2022NatAs...6..141B}, but also in ionized carbon and metastable helium (see Sections \ref{escape_metals} and \ref{escape_He}).

Similar escape signatures in other Neptunes have been observed in Ly$\alpha$, and each of them stand out for a particular reason. GJ~3470~b was observed with \emph{HST} in the Panchromatic Comparative Exoplanetology Treasury (PanCET) program with both the STIS and COS spectrographs \citep{2018A&A...620A.147B, Bourrier21}, yielding a signal of $35\% \pm 7\%$ in the blue wing, which is also explained by a large H exosphere similar to Gl~436~b. A key difference with GJ~3470~b is that it displays an in-transit excess absorption in the Ly$\alpha$ red wing as well, indicating the presence of material inflowing into the star. To this day, the exact physical mechanism behind this inflow remains a mystery, but \citet{2018A&A...620A.147B} tentatively suggests that it could be caused by an elongated layer of dense atomic H extending beyond the Roche lobe. Using COS observations, \citet{DosSantos19} detected a similar, but episodic absorption in the red wing of Gl~436 during one of the observed transits of Gl~436~b in the PanCET program.

In this context, the tentative detection of exopsheric H in K2-18~b by \citet{2020A&A...634L...4D} stands out because this mini-Neptune is not, by any means, a hot exoplanet. Since it orbits an M dwarf with a period of approximately 30 days, K2-18~b is in fact a temperate world. The authors conclude that, due to how faint the host star is in the far-UV, more observations are necessary to confirm the detection. That notwithstanding, a primordial atmosphere of only a few percent mixing ratio of H can lead to temperatures in the upper atmosphere as high as $10\,000$~K, even in a temperate planet \citep{1972JAtS...29..214G}. At these conditions, the kinetic energy of particles in the upper atmosphere exceed the gravitational potential of the planet, leading to a rapid atmospheric expansion and consequent escape. Furthermore, a recent study provided further support that planets at amenable levels of irradiation can sustain a large cloud of atomic H detectable during transits \citep{2021arXiv211106094O}, but this hypothesis still requires further observations to be put under test. Another planet with a tentative Ly$\alpha$ detection is 55~Cnc~b \citep{2012A&A...547A..18E}.

The case of the mini-Neptune HD~63433~c stands out for being the youngest transiting exoplanet with atmospheric escape detected in Ly$\alpha$ \citep{2022AJ....163...68Z}. It orbits a G5-type star with an orbital period of 20.5~d. Interestingly, the inner planet in the system, with a period of 7~d, does not display a Ly$\alpha$ signal, again providing support to the hypothesis that exospheric H in highly-irradiated Neptunes ionizes too quickly to be detectable in our observations. Similarly young exoplanets with signatures of evaporation are expected to be important to disentangle the roles of different escape mechanisms, such as photoevaporation and core-powered mass loss \citep[e.g.,][]{2020MNRAS.493..792G, 2021MNRAS.501L..28K}. However, their observations are challenging due stellar activity modulation \citep{2019AJ....157...96R, 2022arXiv220109905R}, and even when detections have been observed, the interpretation can be complicated since their masses are usually not known \citep[see, however, the case of K2-100~b in][]{2019MNRAS.490..698B}.

\subsection{Non-detections}\label{H_nondet}

There are several reasons why atmospheric escape of H can remain undetected, even for planets that are expected to be evaporating. In Ly$\alpha$, these reasons boil down to: (i) ISM absorption, which absorbs the flux at Doppler velocities where the absorption was supposed to take place; and ii) The host star luminosity yields a low signal-to-noise ratio, which is the usual suspect for M dwarfs. For H$\alpha$, the most likely limitation is the amount of ionized H in the atmosphere, which may not be high enough to produce a detectable signal.

Non-detections of atmospheric escape are severely under-reported, even though they can be just as informative as secure detections. In Table \ref{H_non_detections} we compile a list of non-detections of H escape that have been reported in refereed publications. 

\begin{table}[h]
\caption{List of non-detections of H escape reported in the literature.}
\label{H_non_detections}
\centering
\begin{tabular}{ccl}
\hline
Planet name & Obs. method & Reference \\
\hline
HD~147506~b & H$\alpha$ & \citet{2012ApJ...751...86J} \\
HD~149026~b & H$\alpha$ & \citet{2012ApJ...751...86J} \\ 
HAT-P-32~b & H$\alpha$ & \citet{Mallonn16} \\
KELT-3~b & H$\alpha$ & \citet{2017AJ....153...81C} \\ 
Gl~436~b & H$\alpha$ & \citet{2017AJ....153...81C} \\ 
TRAPPIST-1 system & Ly$\alpha$ & \citet{2017AJ....154..121B} \\
Kepler-444 system & Ly$\alpha$ & \citet{Bourrier17d} \\ 
HD~97658~b & Ly$\alpha$ & \citet{Bourrier17c} \\
55~Cnc~e & Ly$\alpha$ & \citet{Bourrier18} \\
GJ~1132~b & Ly$\alpha$  &\citet{2019AJ....158...50W} \\ 
$\pi$~Men~c & Ly$\alpha$ & \citet{2020ApJ...888L..21G} \\
WASP-29~b & Ly$\alpha$ & \citet{DosSantos2021_WASP-29} \\
K2-25~b & Ly$\alpha$ & \citet{2021AJ....162..116R} \\
GJ~9827~b \& d & Ly$\alpha$ and H$\alpha$& \citet{2021AJ....161..136C} \\
HD~63433~b &Ly$\alpha$ & \citet{2022AJ....163...68Z} \\ 
\hline
\end{tabular}
\end{table}



\section{Hydrodynamic escape of metals observed in the UV}\label{escape_metals}

Other signatures of escape can be observed in UV wavelengths, among them the metal lines of carbon (C), nitrogen (N), oxygen (O), silicon (Si), magnesium (Mg), sulfur (S) and iron (Fe). Since these species are much heavier than H and He, they can only be lifted to the upper atmosphere when the escape is not selective; in other words, metals can only escape when the outflow is in a hydrodynamic regime. Similarly to Ly$\alpha$, these metal lines are present in emission in stars of types between F and M. The advantage of observing these lines is that they do not have ISM absorption, or it is not as dramatic as in Ly$\alpha$. The disadvantage is that metal lines are intrinsically weaker than Ly$\alpha$, which means the detector will register lower count rates, yielding lower signal-to-noise ratios. For this reason, most of the detections of escaping metals have been obtained for hot Jupiters, where the signatures are stronger.

\subsection{Hot Jupiters}

\citet{2004ApJ...604L..69V} first reported on the detection of O\,I and C\,II in the upper atmosphere of HD~209458~b using \emph{HST}/STIS. According to the authors, the high velocity disperson and depth of the in-transit absorption suggests that the escaping metals are outflowing at supersonic velocities above the Roche lobe, an effect also known as geometric blow-off \citep[][]{2004A&A...418L...1L}. In this Roche-lobe filling regime, the mass loss rates of hot exoplanets can be enhanced significantly; in the case of HD~209458~b, \citet{2007A&A...472..329E} found this factor to be in the order of 50\%. Using observations with the COS spectrograph, \citet{2010ApJ...717.1291L} reported on detections of C\,II and Si\,III in HD~209458~b, which is in conflict with the non-detection of Si\,III in \citet{2004ApJ...604L..69V}; however, these COS detections were later contested \citep{2015ApJ...804..116B}. Observations in the near-UV have yielded additional evidence for hydrodynamic escape in this planet associated with the presence of Mg\,I \citep{2013A&A...560A..54V}. According to the authors, the Mg feature probes the thermosphere and the exobase, precisely where the escape takes place; however, they also detect a tentative signal of a Mg comet-like tail in the exosphere of the planet. Finally, \citet{2010ApJ...722L..75S} discussed a tentative detection of Si\,IV in the limb-brightened transit of HD~209458~b.

The archetypal hot Jupiter HD~189733~b was also among the early discoveries of escaping metals \citep{2013A&A...553A..52B}. Despite a significant stellar variability, the transit observations obtained with COS indicated the presence of O\,I and a possible early ingress associated with C\,II. HD~189733~b has since been observed again in the PanCET program, and the analysis of that dataset is currently under way.

Another category of planets that have become a testbed for atmospheric escape is that of the ultrahot Jupiters (UHJ), namely those that orbit closely to stars of type F or earlier. Because they orbit more massive stars, their escaping signatures are frequently detected in a regime of geometric blow-off. WASP-12~b was the first UHJ to have a detection of escaping metals. Using the COS spectrograph, \citet{2010ApJ...714L.222F} reported on excess in-transit absorption signatures in the core of the Mg\,II resonant lines at moderate significance, and on significantly enhanced transit depths measure in wide-band NUV light curves. The wide-band excess absorption are attributed to a collection of different absorbing metals in the exosphere of WASP-12~b. 

\begin{figure}[h]
\begin{center}
\includegraphics[width=0.7\textwidth]{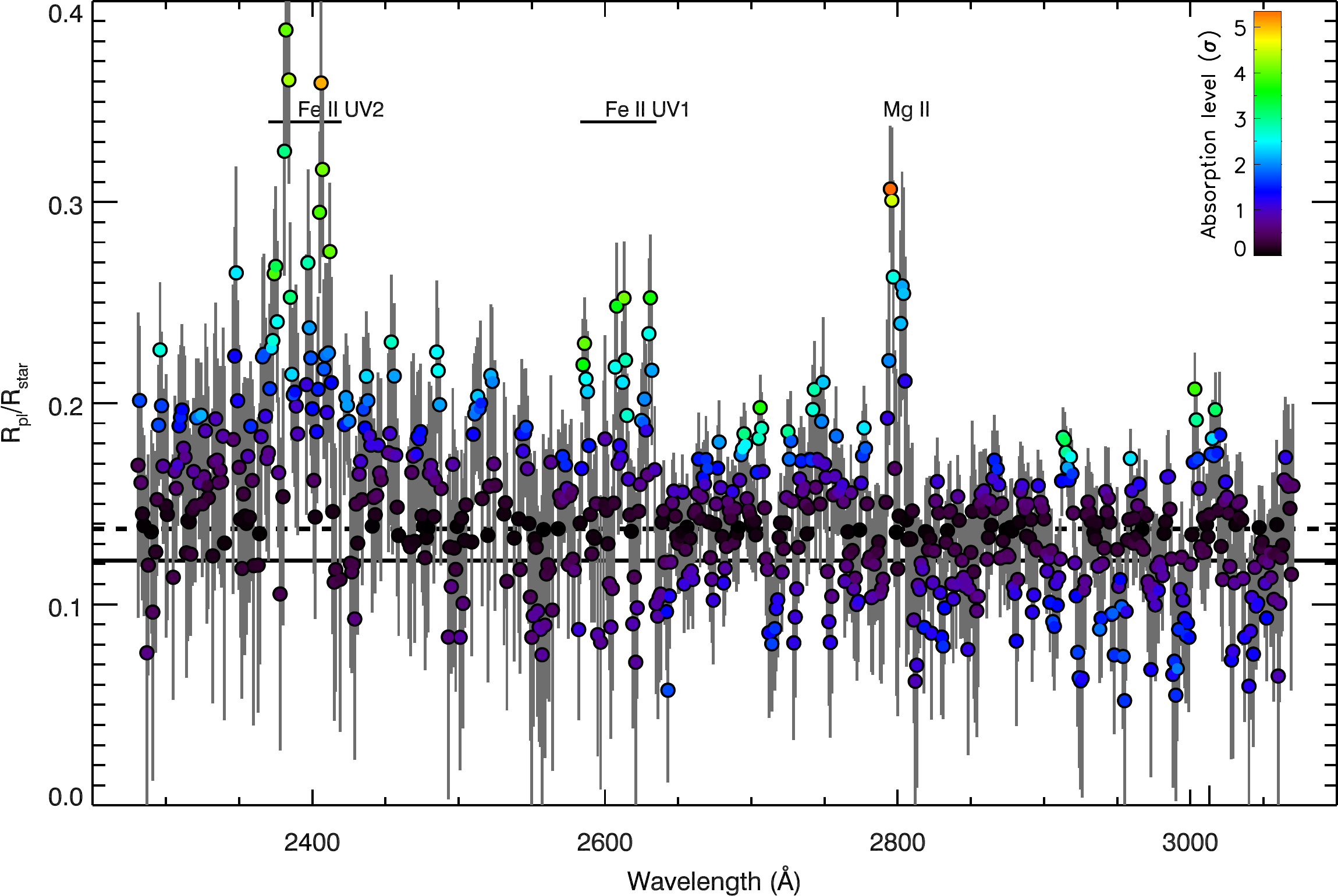} 
\caption{Transmission spectrum of WASP-121~b in the near-UV with detections of Mg\,II and Fe\,II \citep{2019AJ....158...91S}. Reproduced with the permission of AAS Journals.} 
\label{fig:W121b_metals}
\end{center}
\end{figure}

The latest UHJ in which escaping metals have been detected is WASP-121~b \citep{2019AJ....158...91S}. Based on STIS observations, the authors find evidence of Mg\,II and Fe\,II ions filling the Roche-lobe of the planet (see Fig. \ref{fig:W121b_metals}), and deeper broadband NUV light curves compared to optical wavelengths.

\subsection{Neptunes and sub-Neptunes}

To date, only two sub-Jovian worlds have been shown to display signatures of escaping metals, both of them obtained with \emph{HST}/COS. The warm Neptune HAT-P-11~b has an excess in-transit absorption of $15\% \pm 4\%$ in the blue wing of the ground-state C\,II line at 133.45~nm, as well as a post-transit tail absorption of $12.5\% \pm 4\%$ \citep{2022NatAs...6..141B}. The authors argue that this signal is consistent with the planet's atmosphere having a sub-solar metallicity and an extended magnetotail.

The second, and perhaps most intriguing detection is that of the super-Earth $\pi$~Men~c \citep{2021ApJ...907L..36G}. Benefitting from the brightness of the host star, the authors reported on an in-transit absorption of $3.9\% \pm 1.1\%$ in the blue wing of the excited-state C\,II line at 1335~\AA. Based on this single-transit observation, the authors concluded that $\pi$~Men~c possesses a thick atmosphere with more than $50\%$ heavy volatiles in mass fraction, and that the escaping C fills the Roche lobe of the planet.

Although STIS observations hinted at a tentative detection of Si\,III in Gl~436~b \citep{2017A&A...605L...7L}, an ensemble of COS data was later used to show that the observed signal was not present \citep{Loyd17}, and that the STIS data was likely contaminated by stellar activity modulation \citep{DosSantos19}.

\subsection{Non-detections}

Similarly to Ly$\alpha$, observations of metals in the UV suffer from the low signal-to-noise ratios, and this is probably the main limitation for transmission spectroscopy in these wavelengths. As seen in the case of $\pi$~Men~c \citep{2021ApJ...907L..36G}, some of the in-transit signals we are looking for are in the order of only a few percent, which requires high levels of contrast in order to be detected. Additionally, as shown by \citet{DosSantos19}, stellar activity can also pose as a false positive. We list the non-detections of escaping metals reported in the literature in Table \ref{metal_non_detections}.

\begin{table}[h]
\caption{List of non-detections of escaping metals reported in the literature.}
\label{metal_non_detections}
\centering
\begin{tabular}{ccl}
\hline
Planet name & Instrument & Reference \\
\hline
WASP-13~b & \emph{HST}/COS & \citet{2015ApJ...815..118F} \\
55~Cnc~e & Ground-based spectrographs & \citet{Ridden16} \\
Gl~436~b & \emph{HST}/STIS \& COS & \citet{Loyd17, DosSantos19} \\ 
WASP-18~b & \emph{HST}/COS & \citet{2018AJ....155..113F} \\
WASP-29~b & \emph{HST}/COS & \citet{DosSantos2021_WASP-29} \\
HD~189733~b & XMM-Newton optical monitor & \citet{2021MNRAS.506.2453K} \\
GJ~3470~b & \emph{HST}/COS & \citet{Bourrier21} \\
\hline
\end{tabular}
\end{table}

\section{Metastable He spectroscopy in the near-infrared}\label{escape_He}

Classically, the near-infrared helium (He) triplet located at 1.083~$\mu$m has been used to probe the chromosphere and transition region of cool stars \citep[e.g.,][]{1997ApJ...489..375A}. The presence of He in the upper atmospheres of exoplanets was originally predicted by the theoretical models of \citet{2000ApJ...537..916S}, but early observations of HD~209458~b were unable to detect a signal \citep{Moutou03}. Several years later, \citet{2018ApJ...855L..11O} predicted that escaping He could produce signals as deep as 6\% in the core of the triplet of HD~209458~b, which could be detectable at high spectral resolution. 

Neutral He atoms can exist in two states: singlet (1$^1$S, electrons with anti-parallel spin) or triplet (2$^3$S, electrons with parallel spin). Since the radiative decay of triplet He into singlet state is relatively long, the former is also known as a metastable state. The formation of this line depends on the balance of rates that either populate or de-populate the triplet state: recombination, collisional excitation and de-excitation, charge exchange, and photoionization. According to \citet{2019ApJ...881..133O}, planets orbiting late-type and active stars tend to display prominent in-transit He absorption due to their favorably high levels of extreme UV flux. \citet{2022MNRAS.512.1751P} further proposed that metastable He absorption also has a dependence on the iron abundance in the corona of stellar hosts, since most of the extreme-UV flux comes from coronal iron emission lines in cool stars. As we shall see shortly, this trend has mostly been held in our observations. 

The most important advantage of observing metastable He is that this technique does not necessarilly require a space telescope, and can be observed from the ground. In fact, ground-based facilities can perform experiments at much higher spectral resolutions that those achieved from space. In this regime, the Doppler anomaly of the planet can be resolved during the transit \citep[e.g.,][]{2015A&A...577A..62W}, which helps in discerning if the signal is of planetary nature or stellar. As opposed to Ly$\alpha$ observations, the in-transit absorption is seen in the core and wings of the He triplet, which means we are not only probing the accelerated particles well above the exobase, but also the outflowing gas near the thermosphere. This allows us to use simpler, one-dimensional models to interpret the observations \citep[e.g.,][]{2018ApJ...855L..11O, 2020A&A...636A..13L, 2022A&A...659A..62D, 2022arXiv220903677L} and extract more precise mass loss rates than those determined from Ly$\alpha$ data \citep[e.g.,][]{2013A&A...557A.124B}. The disadvantage of ground-based He spectroscopy is that, due to spectral normalization, information about the planetary continuum absorption is lost, but since the signals are relatively deep, the impact of this limitation is not of great importance. Other disadvantages include telluric contamination and lower sensitivities than space telescopes.

\subsection{Hot Jupiters}

For a change, the first discovery of metastable He in exoplanet was not in HD~209458~b, but rather the hot Jupiter WASP-107~b \citep{2018Natur.557...68S}. In this study, the authors observed a single transit with \emph{HST} and the Wide-Field Camera 3 (WFC3) instrument and measured a transit depth of $0.049\% \pm 0.011\%$ in a low-resolution bandpass of 98~\AA. Later, this feature would be observed again from the ground and at high spectral resolution with the CARMENES spectrograph installed on the 3.5~m telescope at the Calar Alto Observatory \citep{2019A&A...623A..58A} and with the Keck II/NIRSPEC spectrograph \citep{2020AJ....159..115K}. Recently, \citet{2021AJ....162..284S} reported on the observation of the He tail that trails WASP-107~b, also detected with the NIRSPEC instrument.

Several other hot Jupiters have since been observed to be evaporating and exhibit in-transit He absorption. The CARMENES spectrograph has been particularly productive, yielding detections for HD~189733~b \citep[with variability; ][]{2018A&A...620A..97S}, WASP-69~b \citep{2018Sci...362.1388N}, HD~209458~b \citep{2019A&A...629A.110A}, HAT-P-32~b \citep{2022A&A...657A...6C} and a tentative detection for the UHJ WASP-76~b \citep{2021A&A...654A.163C}. Another productive instrument for He spectroscopy in hot Jupiters has been the NIRSPEC spectrograph, which was responsible for detections in HD~198733~b \citep{2022arXiv220402985Z}, WASP-52~b and a tentative signal for WASP-177~b \citep{2022AJ....164...24K}. Using the GIANO spectrograph installed on the Telescopio Nazionale Galileo (TNG), \citet{2020A&A...639A..49G} reproduced the He signature of HD~189733~b.

Following up on the increasing interest in atmospheric escape in exoplanets, \citet{2020AJ....159..278V} presented the first results detections of He using a new method: ultra-narrowband photometry of the He triplet with the Wide-field Infrared Camera (WIRC) installed on the 200-inch Hale Telescope at Palomar Observatory. They use a custom-made filter centered on 1083.3~nm in vacuum, with an FWHM of 0.635~nm, and a maximum transmission of 95.6\%. Naturally, since measurements are performed in photometry, the in-transit absorption is not spectrally resolved, so the results do not encode information about velocities. However, the instrument has demonstrated a significant productivity, yielding detections for the hot Jupiters WASP-69~b \citep{2020AJ....159..278V}, HAT-P-18~b \citep{2021ApJ...909L..10P}, and tentative detections for WASP-52~b and NGTS-5~b \citep{2022arXiv220411865V}. This tentative observation for WASP-52~b, with an in-transit depth of $0.29\% \pm 0.13\%$ in the filter's bandpass, is in slight tension with the firm detection reported in \citet{2022AJ....164...24K}, for which an in-transit depth of $0.66\% \pm 0.14\%$ was measured.

\subsection{Neptunes and sub-Neptunes}

\begin{figure}[h]
\begin{center}
\includegraphics[width=0.7\textwidth]{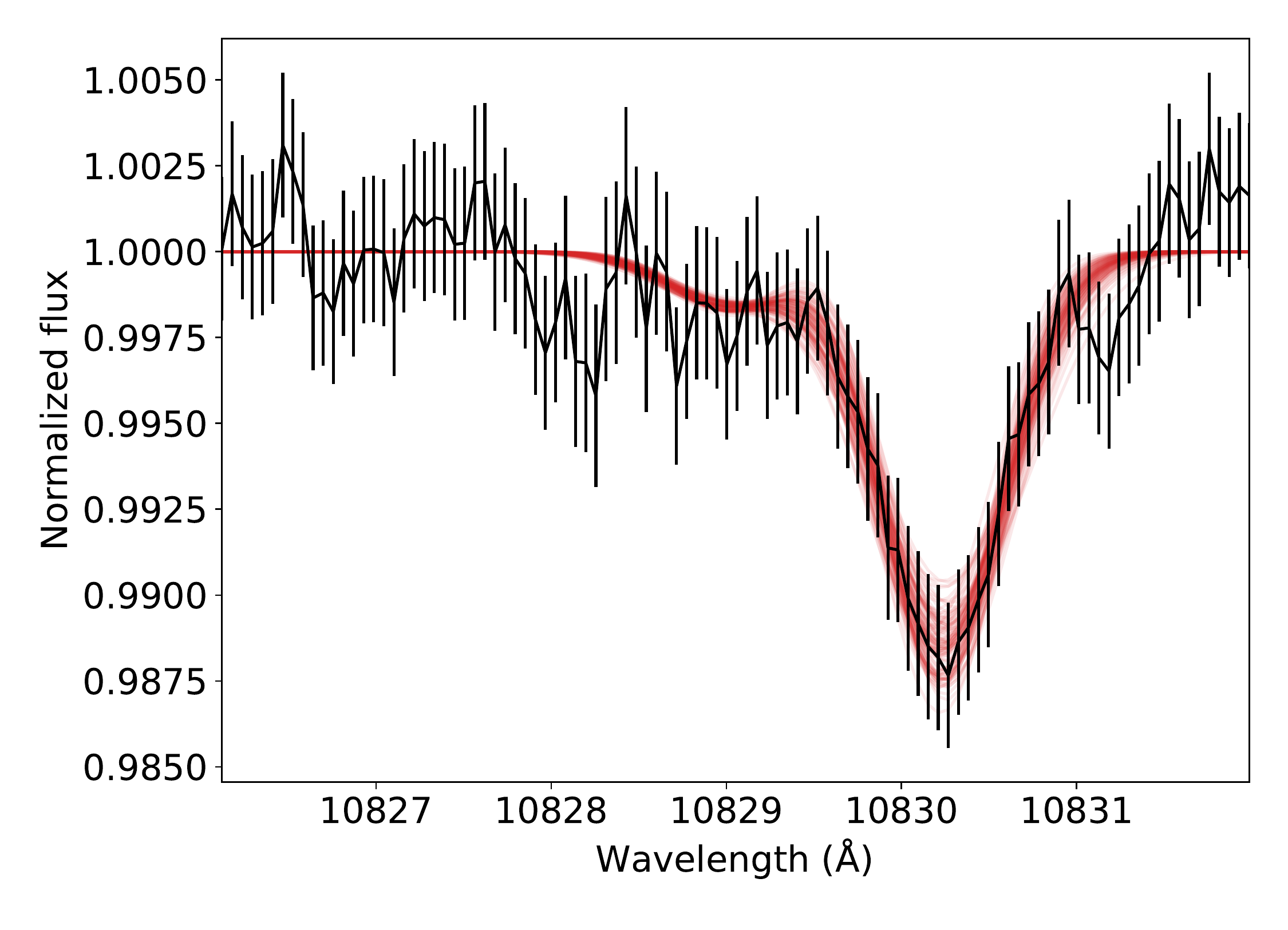} 
\caption{Transmission spectrum of HAT-P-11~b near the metastable He triplet observed with the CARMENES spectrograph \citep[black symbols;][]{2018Sci...362.1384A} and a family of transmission spectra simulations that were fit to the data based on an isothermal Parker-wind model \citep[red curves;][]{2022A&A...659A..62D}.} 
\label{fig:hp11b_He}
\end{center}
\end{figure}

The first reports of He observations in transiting hot Jupiters were concomitant with the discoveries in sub-Jovian worlds. Similar to Ly$\alpha$ results we listed in Sect. \ref{escape_H}, Neptunes can also display deep in-transit signals that sometimes rival those of their larger counterparts. Using an \emph{HST}/WFC3 archival dataset, \citet{2018ApJ...868L..34M} demonstrated that the warm Neptune HAT-P-11~b has a transit depth of $\sim 0.355\%$ in a 49~\AA-wide channel centered in the He triplet. This study was simultaneous to that of \citet{2018Sci...362.1384A}, who reported a detection of He in HAT-P-11~b obtained with the CARMENES spectrograph. At high spectral resolution, this signature is resolved with an average depth of $1.08\% \pm 0.05\%$ (see Fig. \ref{fig:hp11b_He}).

Along with HAT-P-11~b, another warm Neptune to have both Ly$\alpha$ and He detections is GJ~3470~b \citep{2020ApJ...894...97N}, the latter obtained with the Habitable Zone Planet Finder (HPF) spectrograph installed on the Hobby-Eberly Telescope (HET). This signal was also measured with CARMENES \citep{2020A&A...638A..61P}, and a large mass loss rate of $\sim 10^{11}$~g\,s$^{-1}$ was inferred. Based on this observation, \citet{2021A&A...648L...7L} concludes that GJ~3470~b is in a photon-limited escape regime, where the mass loss rate is limited by the incident flux of ionizing photons \citep{2016ApJ...816...34O}.

Using the NIRSPEC spectrograph, \citet{2020AJ....160..258K} concluded that the sub-Neptune GJ~1214~b does not have a detectable He signature. However, the observation of one transit with the CARMENES spectrograph reported by \citet{2022A&A...659A..55O} yielded a tentative detection. According to \citeauthor{2022A&A...659A..55O}, stellar activity alone cannot have caused a false-positive, and argue that telluric contamination is the probable culprit of the non-detection observed by \citeauthor{2020AJ....160..258K}. This argument was further contested by \citet{2022arXiv220903502S}, who observed an additional NIRSPEC transit in an epoch of minimal telluric contamination and still obtained a non-detection. 

The Palomar/WIRC narrowband photometry has proven to be precise enough to detect He outflows in Neptunes as well, with a firm detection for HAT-P-26~b \citep{2022arXiv220411865V} and a tentative detection for the young sub-Neptune V1298~Tau~d \citep{2021AJ....162..222V}. Young sub-Neptunes are target of strong importance for atmospheric escape observations because we think it is in their youth that most of photoevaporation takes place \citep[e.g.,][]{2013ApJ...775..105O}. Using the NIRSPEC spectrograph, \citet{2022AJ....163...67Z, 2022arXiv220713099Z} reported on the first discoveries of atmospheric escape in the young mini-Neptunes HD~73583~b, TOI-1430~b, TOI-2076~b and TOI-1683~b.



\subsection{Non-detections}

Many more planets than those listed in this review have been observed in He spectroscopy (priv. comm.), and the data analyses are currently under way. We list all the reported He non-detections in Table \ref{He_non_detections}, where we also include those that have been positively detected using different instruments or analyses (these cases are marked with the symbol $^\dagger$).

\begin{table}[h]
\caption{List of non-detections of metastable He reported in the literature.}
\label{He_non_detections}
\centering
\begin{tabular}{ccl}
\hline
Planet name & Instrument & Reference \\
\hline
HD~209458~b$^\dagger$ & VLT/ISAAC & \citet{Moutou03} \\
Gl~436~b & CAO~3.5m/CARMENES & \citet{2018Sci...362.1388N} \\
KELT-9~b & CAO~3.5m/CARMENES & \citet{2018Sci...362.1388N} \\
WASP-12~b & \emph{HST}/WFC3 & \citet{2018RNAAS...2...44K} \\ 
WASP-52~b$^\dagger$ & Palomar/WIRC & \citet{2020AJ....159..278V} \\  
K2-100~b & Subaru/IRD & \citet{2020MNRAS.495..650G} \\
WASP-127~b & Gemini/Phoenix & \citet{DosSantos20} \\
AU~Mic~b & Subaru/IRD \& Keck II/NIRSPEC & \citet{2020ApJ...899L..13H} \\ 
GJ~1214~b$^\dagger$ & Keck II/NIRSPEC & \citet{2020AJ....160..258K, 2022arXiv220903502S} \\
HD~97658~b & Keck II/NIRSPEC & \citet{2020AJ....160..258K} \\
55~Cnc~e & Keck II/NIRSPEC & \citet{2021AJ....161..181Z} \\
TRAPPIST-1 system & Subaru/IRD \& HET/HPF & \citet{2021AJ....162...82K} \\
K2-136~c & Subaru/IRD & \citet{2021RNAAS...5..238G} \\
V1298~Tau~b \& c & Palomar/WIRC & \citet{2021AJ....162..222V} \\
WASP-80~b & TNG/GIANO \& Palomar/WIRC & \citet{Fossati22, 2022arXiv220411865V} \\
GJ~9827~d & Keck II/NIRSPEC & \citet{2020AJ....160..258K} \\
$\tau$~Boo~b (in emission) & CAO~3.5m/CARMENES & \citet{Zhang20} \\
GJ~9827~b \& d & CAO~3.5m/CARMENES & \citet{2021AJ....161..136C} \\
HD 63433 system & Keck II/NIRSPEC & \citet{2022AJ....163...68Z} \\
WASP-177~b$^\dagger$ & Palomar/WIRC & \citet{2022arXiv220411865V} \\
\hline
\multicolumn{3}{l}{Notes: The $^\dagger$ symbol denotes planets with alternate results that yielded a detection.} \\
\end{tabular}
\end{table}

\section{Conclusions and future perspectives}\label{future}

Atmospheric escape has been studied in the Solar System since the beginning of the 20th Century. However, observations of upper atmospheres in hot exoplanets in the last two decades have advanced our understanding about the physics of evaporation by leaps and bounds. To date, we have observed escape in 28 exoplanets, including tentative detections (see a complete list in Table \ref{list_detections}). These worlds have sizes varying from Jupiter-size to mini-Neptunes, and irradiation levels ranging from the most extremely-irradiated planet known (KELT-9~b) to Earth-like bolometric fluxes (K2-18~b).

\begin{table}[h]
\caption{List of planets with detections of atmospheric escape.}
\label{list_detections}
\centering
\begin{tabular}{cc}
\hline
Planet name & Signature(s) \\
\hline
HD 209458 b & Ly$\alpha$, Balmer-series, metals, He \\
HD 189733 b & Ly$\alpha$, Balmer-series, metals, He \\
KELT-9 b  & Balmer-series \\
WASP-12 b  & Balmer-series, metals \\
KELT-20 b  & Balmer-series \\
WASP-52 b & Balmer-series, He \\
WASP-33 b  & Balmer-series \\
WASP-121 b & Balmer-series, metals \\
Gl 436 b & Ly$\alpha$ \\
GJ 3470 b & Ly$\alpha$, He \\
K2-18 b & Ly$\alpha$ (tentative) \\
HD 63433 c & Ly$\alpha$ \\
HAT-P-11 b & Ly$\alpha$, C\,II, He \\
$\pi$ Men c & C\,II \\
WASP-107 b & He \\
WASP-69 b & He \\
HAT-P-32 b & He \\
WASP-76 b & He (tentative) \\
WASP-177 b & He (tentative) \\
GJ~1214~b & He (tentative) \\
HAT-P-18 b & He \\
NGTS-5 b &  He (tentative) \\
HAT-P-26 b & He  \\
V1298 Tau d &  He (tentative) \\
HD 73583 b & He \\
TOI-1430 b & He \\
TOI-2076 b & He \\
TOI-1683 b & He \\
\hline
\end{tabular}
\end{table}

Our observational efforts have shown that, so far, metastable He spectroscopy is the most productive avenue to observe escape in hot exoplanets with a H-dominated atmosphere orbiting active stars. Ly$\alpha$ observations, on the other hand, seem to yield detections for planets in relatively milder irradiation conditions. According to \citet{2021arXiv211106094O}, the reason for that is due to lack of observable flux in the core of the Ly$\alpha$ line, which means that we have access only to signatures that occur at high Doppler velocities. In order for H atoms to achieve these high velocities, they need to stay neutral for a long time and produce a detectable exospheric tail. For the cases where H ionizes too quickly in the exosphere, it is thus recommended to observe in the Balmer series lines. Some hot Jupiters, like HD~209458~b and HD~189733~b, have an optimal set of parameters that allows the detection of Ly$\alpha$, He, H$\alpha$, and metals. Some Neptunes, like GJ~3470~b and HAT-P-11~b, also possess an optimal set of parameters that enables the observation of escape in more than one spectral channel. These cases seem to be, however, rare.

Despite these observational efforts, many questions related to the atmospheric evolution of exoplanets remain open. For instance, what are the mechanisms that carve the hot Neptune desert \citep[e.g.,][]{2009MNRAS.396.1012D, 2011ApJ...727L..44S, 2016A&A...589A..75M}? Based on a survey of escaping He in Saturn-sized ho gas giants with Palomar/WIRC, \citet{2022arXiv220411865V} concluded that the upper edge of the Neptune desert is stable against evaporation, with measured escape rates that remove less than 10\% of these planet's masses. This suggests that other, additional mechanisms are necessary to carve the desert, such as a history of migration \citep[e.g.,][]{2018MNRAS.479.5012O, 2021A&A...647A..40A}. More observations and modeling are required to test these hypotheses. Another persistent open question in this field is whether close-in gas giant exoplanets have hydrodynamically unstable thermospheres \citep{2016A&A...585L...2S}, which was originally proposed by \citet{1981Icar...48..150W} and confirmed for only a handful of exoplanets to date. More observations of exospheric metals will help elucidate this puzzle, since they trace hydrodynamic escape directly.

For the future, as we mentioned in Sect. \ref{escape_He}, observing escape in young sub-Neptunes will also be important because it may give us clues about the respective roles of photoevaporation \citep[driven by X-rays and extreme-UV irradiation; e.g.,][]{2003ApJ...598L.121L, 2011ApJ...729L..24V, 2015ApJ...808..173T, 2016MNRAS.460.1300E} and core-powered mass loss \citep[e.g.,][]{2018MNRAS.476..759G, 2019MNRAS.487...24G}. The main challenge in this endeavor is that young stars are active, and the activity poses a problem to measure planetary masses through the radial velocity method and, in addition, can produce false-positive detections of escape \citep[e.g.,][]{DosSantos19}.

With a sample of 28 exoplanets with signals of atmospheric escape, we have by now gathered a sample with which we can begin interpreting at a comparative level. In order to answer some of the open questions described above, we will benefit from carrying out a uniform analysis of this sample with a common theoretical framework \citep[see, e.g.,][]{2016Natur.529...59S}. To cite an example, this approach will enable us to find correlations between measured properties of evaporating exoplanets, such as their bulk density, incoming high-energy flux, and mass-loss rates (as predicted by the energy-limited formulation). Studies that have already begun performing this comparative exoplanetology approach for evaporating exoplanets are \citet{2021A&A...648L...7L} and \citet{2022arXiv220411865V}. 

Finally, with the successful launch and commissioning of \emph{JWST}, we will have yet another instrument capable of observing the metastable He line, and with space-based precision. Although its capabilities for He transmission spectroscopy remain to be tested, it has three instrument configurations that can measure spectra at 1.083~$\mu m$: NIRISS/SOSS (2nd order only), NIRSpec/G140M and NIRSpec/G140H. The downside of \emph{JWST} is that its lower resolution may not be able to spectrally resolve the in-transit absorption, but a more precise instrument could enable us to observe fainter signatures than those accessible from the ground.

\def\apj{{ApJ}}    
\def\nat{{Nature}}    
\def\jgr{{JGR}}    
\def\apjl{{ApJ Letters}}    
\def\aap{{A\&A}}   
\def\mnras{{MNRAS}}
\def\aj{{AJ}}
\def\icarus{{Icarus}}
\def\apss{{Astrophysics and Space Science}}
\let\mnrasl=\mnras


\bibliographystyle{aa}
\bibliography{references}

\begin{thebibliography}{149}
\expandafter\ifx\csname natexlab\endcsname\relax\def\natexlab#1{#1}\fi

\bibitem[{{Allart} {et~al.}(2019){Allart}, {Bourrier}, {Lovis}, {Ehrenreich},
  {Aceituno}, {Guijarro}, {Pepe}, {Sing}, {Spake}, \&
  {Wyttenbach}}]{2019A&A...623A..58A}
{Allart}, R., {Bourrier}, V., {Lovis}, C., {et~al.} 2019, \aap, 623, A58

\bibitem[{{Allart} {et~al.}(2018){Allart}, {Bourrier}, {Lovis}, {Ehrenreich},
  {Spake}, {Wyttenbach}, {Pino}, {Pepe}, {Sing}, \& {Lecavelier des
  Etangs}}]{2018Sci...362.1384A}
{Allart}, R., {Bourrier}, V., {Lovis}, C., {et~al.} 2018, Science, 362, 1384

\bibitem[{{Alonso-Floriano} {et~al.}(2019){Alonso-Floriano}, {Snellen},
  {Czesla}, {Bauer}, {Salz}, {Lamp{\'o}n}, {Lara}, {Nagel},
  {L{\'o}pez-Puertas}, {Nortmann}, {S{\'a}nchez-L{\'o}pez}, {Sanz-Forcada},
  {Caballero}, {Reiners}, {Ribas}, {Quirrenbach}, {Amado}, {Aceituno},
  {Anglada-Escud{\'e}}, {B{\'e}jar}, {Brinkm{\"o}ller}, {Hatzes}, {Henning},
  {Kaminski}, {K{\"u}rster}, {Labarga}, {Montes}, {Pall{\'e}}, {Schmitt}, \&
  {Zapatero Osorio}}]{2019A&A...629A.110A}
{Alonso-Floriano}, F.~J., {Snellen}, I.~A.~G., {Czesla}, S., {et~al.} 2019,
  \aap, 629, A110

\bibitem[{{Andretta} \& {Jones}(1997)}]{1997ApJ...489..375A}
{Andretta}, V. \& {Jones}, H.~P. 1997, \apj, 489, 375

\bibitem[{{Attia} {et~al.}(2021){Attia}, {Bourrier}, {Eggenberger},
  {Mordasini}, {Beust}, \& {Ehrenreich}}]{2021A&A...647A..40A}
{Attia}, O., {Bourrier}, V., {Eggenberger}, P., {et~al.} 2021, \aap, 647, A40

\bibitem[{{Ballester} \& {Ben-Jaffel}(2015)}]{2015ApJ...804..116B}
{Ballester}, G.~E. \& {Ben-Jaffel}, L. 2015, \apj, 804, 116

\bibitem[{{Ballester} {et~al.}(2007){Ballester}, {Sing}, \&
  {Herbert}}]{2007Natur.445..511B}
{Ballester}, G.~E., {Sing}, D.~K., \& {Herbert}, F. 2007, \nat, 445, 511

\bibitem[{{Barrag{\'a}n} {et~al.}(2019){Barrag{\'a}n}, {Aigrain}, {Kubyshkina},
  {Gandolfi}, {Livingston}, {Fridlund}, {Fossati}, {Korth}, {Parviainen},
  {Malavolta}, {Palle}, {Deeg}, {Nowak}, {Rajpaul}, {Zicher}, {Antoniciello},
  {Narita}, {Albrecht}, {Bedin}, {Cabrera}, {Cochran}, {de Leon},
  {Eigm{\"u}ller}, {Fukui}, {Granata}, {Grziwa}, {Guenther}, {Hatzes},
  {Kusakabe}, {Latham}, {Libralato}, {Luque},
  {Monta{\~n}{\'e}s-Rodr{\'\i}guez}, {Murgas}, {Nardiello}, {Pagano}, {Piotto},
  {Persson}, {Redfield}, \& {Tamura}}]{2019MNRAS.490..698B}
{Barrag{\'a}n}, O., {Aigrain}, S., {Kubyshkina}, D., {et~al.} 2019, \mnras,
  490, 698

\bibitem[{{Ben-Jaffel} \& {Ballester}(2013)}]{2013A&A...553A..52B}
{Ben-Jaffel}, L. \& {Ballester}, G.~E. 2013, \aap, 553, A52

\bibitem[{{Ben-Jaffel} {et~al.}(2022){Ben-Jaffel}, {Ballester}, {Garc{\'\i}a
  Mu{\~n}oz}, {Lavvas}, {Sing}, {Sanz-Forcada}, {Cohen}, {Kataria}, {Henry},
  {Buchhave}, {Mikal-Evans}, {Wakeford}, \&
  {L{\'o}pez-Morales}}]{2022NatAs...6..141B}
{Ben-Jaffel}, L., {Ballester}, G.~E., {Garc{\'\i}a Mu{\~n}oz}, A., {et~al.}
  2022, Nature Astronomy, 6, 141

\bibitem[{{Bourrier} {et~al.}(2017{\natexlab{a}}){Bourrier}, {de Wit},
  {Bolmont}, {Stamenkovi{\'c}}, {Wheatley}, {Burgasser}, {Delrez}, {Demory},
  {Ehrenreich}, {Gillon}, {Jehin}, {Leconte}, {Lederer}, {Lewis}, {Triaud}, \&
  {Van Grootel}}]{2017AJ....154..121B}
{Bourrier}, V., {de Wit}, J., {Bolmont}, E., {et~al.} 2017{\natexlab{a}}, \aj,
  154, 121

\bibitem[{{Bourrier} {et~al.}(2021){Bourrier}, {Dos Santos}, {Sanz-Forcada},
  {Garc{\'\i}a Mu{\~n}oz}, {Henry}, {Lavvas}, {Lecavelier},
  {L{\'o}pez-Morales}, {Mikal-Evans}, {Sing}, {Wakeford}, \&
  {Ehrenreich}}]{Bourrier21}
{Bourrier}, V., {Dos Santos}, L.~A., {Sanz-Forcada}, J., {et~al.} 2021, \aap,
  650, A73

\bibitem[{{Bourrier} {et~al.}(2017{\natexlab{b}}){Bourrier}, {Ehrenreich},
  {Allart}, {Wyttenbach}, {Semaan}, {Astudillo-Defru}, {Gracia-Bern{\'a}},
  {Lovis}, {Pepe}, {Thomas}, \& {Udry}}]{Bourrier17d}
{Bourrier}, V., {Ehrenreich}, D., {Allart}, R., {et~al.} 2017{\natexlab{b}},
  \aap, 602, A106

\bibitem[{{Bourrier} {et~al.}(2017{\natexlab{c}}){Bourrier}, {Ehrenreich},
  {King}, {Lecavelier des Etangs}, {Wheatley}, {Vidal-Madjar}, {Pepe}, \&
  {Udry}}]{Bourrier17c}
{Bourrier}, V., {Ehrenreich}, D., {King}, G., {et~al.} 2017{\natexlab{c}},
  \aap, 597, A26

\bibitem[{{Bourrier} {et~al.}(2015){Bourrier}, {Ehrenreich}, \& {Lecavelier des
  Etangs}}]{2015A&A...582A..65B}
{Bourrier}, V., {Ehrenreich}, D., \& {Lecavelier des Etangs}, A. 2015, \aap,
  582, A65

\bibitem[{{Bourrier} {et~al.}(2018{\natexlab{a}}){Bourrier}, {Ehrenreich},
  {Lecavelier des Etangs}, {Louden}, {Wheatley}, {Wyttenbach}, {Vidal-Madjar},
  {Lavie}, {Pepe}, \& {Udry}}]{Bourrier18}
{Bourrier}, V., {Ehrenreich}, D., {Lecavelier des Etangs}, A., {et~al.}
  2018{\natexlab{a}}, \aap, 615, A117

\bibitem[{{Bourrier} \& {Lecavelier des Etangs}(2013)}]{2013A&A...557A.124B}
{Bourrier}, V. \& {Lecavelier des Etangs}, A. 2013, \aap, 557, A124

\bibitem[{{Bourrier} {et~al.}(2013){Bourrier}, {Lecavelier des Etangs},
  {Dupuy}, {Ehrenreich}, {Vidal-Madjar}, {H{\'e}brard}, {Ballester},
  {D{\'e}sert}, {Ferlet}, {Sing}, \& {Wheatley}}]{2013A&A...551A..63B}
{Bourrier}, V., {Lecavelier des Etangs}, A., {Dupuy}, H., {et~al.} 2013, \aap,
  551, A63

\bibitem[{{Bourrier} {et~al.}(2018{\natexlab{b}}){Bourrier}, {Lecavelier des
  Etangs}, {Ehrenreich}, {Sanz-Forcada}, {Allart}, {Ballester}, {Buchhave},
  {Cohen}, {Deming}, {Evans}, {Garc{\'{\i}}a Mu{\~n}oz}, {Henry}, {Kataria},
  {Lavvas}, {Lewis}, {L{\'o}pez-Morales}, {Marley}, {Sing}, \&
  {Wakeford}}]{2018A&A...620A.147B}
{Bourrier}, V., {Lecavelier des Etangs}, A., {Ehrenreich}, D., {et~al.}
  2018{\natexlab{b}}, \aap, 620, A147

\bibitem[{{Bourrier} {et~al.}(2016){Bourrier}, {Lecavelier des Etangs},
  {Ehrenreich}, {Tanaka}, \& {Vidotto}}]{2016A&A...591A.121B}
{Bourrier}, V., {Lecavelier des Etangs}, A., {Ehrenreich}, D., {Tanaka}, Y.~A.,
  \& {Vidotto}, A.~A. 2016, \aap, 591, A121

\bibitem[{{Bourrier} {et~al.}(2020){Bourrier}, {Wheatley}, {Lecavelier des
  Etangs}, {King}, {Louden}, {Ehrenreich}, {Fares}, {Helling}, {Llama},
  {Jardine}, \& {Vidotto}}]{2020MNRAS.493..559B}
{Bourrier}, V., {Wheatley}, P.~J., {Lecavelier des Etangs}, A., {et~al.} 2020,
  \mnras, 493, 559

\bibitem[{{Carleo} {et~al.}(2021){Carleo}, {Youngblood}, {Redfield}, {Casasayas
  Barris}, {Ayres}, {Vannier}, {Fossati}, {Palle}, {Livingston}, {Lanza},
  {Niraula}, {Alvarado-G{\'o}mez}, {Chen}, {Gandolfi}, {Guenther}, {Linsky},
  {Nagel}, {Narita}, {Nortmann}, {Shkolnik}, \&
  {Stangret}}]{2021AJ....161..136C}
{Carleo}, I., {Youngblood}, A., {Redfield}, S., {et~al.} 2021, \aj, 161, 136

\bibitem[{{Carolan} {et~al.}(2021){Carolan}, {Vidotto}, {Villarreal D'Angelo},
  \& {Hazra}}]{2021MNRAS.500.3382C}
{Carolan}, S., {Vidotto}, A.~A., {Villarreal D'Angelo}, C., \& {Hazra}, G.
  2021, \mnras, 500, 3382

\bibitem[{{Casasayas-Barris} {et~al.}(2021){Casasayas-Barris}, {Orell-Miquel},
  {Stangret}, {Nortmann}, {Yan}, {Oshagh}, {Palle}, {Sanz-Forcada},
  {L{\'o}pez-Puertas}, {Nagel}, {Luque}, {Morello}, {Snellen}, {Zechmeister},
  {Quirrenbach}, {Caballero}, {Ribas}, {Reiners}, {Amado}, {Bergond}, {Czesla},
  {Henning}, {Khalafinejad}, {Molaverdikhani}, {Montes}, {Perger},
  {S{\'a}nchez-L{\'o}pez}, \& {Sedaghati}}]{2021A&A...654A.163C}
{Casasayas-Barris}, N., {Orell-Miquel}, J., {Stangret}, M., {et~al.} 2021,
  \aap, 654, A163

\bibitem[{{Casasayas-Barris} {et~al.}(2018){Casasayas-Barris}, {Pall{\'e}},
  {Yan}, {Chen}, {Albrecht}, {Nortmann}, {Van Eylen}, {Snellen}, {Talens},
  {Gonz{\'a}lez Hern{\'a}ndez}, {Rebolo}, \& {Otten}}]{2018A&A...616A.151C}
{Casasayas-Barris}, N., {Pall{\'e}}, E., {Yan}, F., {et~al.} 2018, \aap, 616,
  A151

\bibitem[{{Cauley} {et~al.}(2017{\natexlab{a}}){Cauley}, {Redfield}, \&
  {Jensen}}]{2017AJ....153..217C}
{Cauley}, P.~W., {Redfield}, S., \& {Jensen}, A.~G. 2017{\natexlab{a}}, \aj,
  153, 217

\bibitem[{{Cauley} {et~al.}(2017{\natexlab{b}}){Cauley}, {Redfield}, \&
  {Jensen}}]{2017AJ....153...81C}
{Cauley}, P.~W., {Redfield}, S., \& {Jensen}, A.~G. 2017{\natexlab{b}}, \aj,
  153, 81

\bibitem[{{Cauley} {et~al.}(2019){Cauley}, {Shkolnik}, {Ilyin}, {Strassmeier},
  {Redfield}, \& {Jensen}}]{2019AJ....157...69C}
{Cauley}, P.~W., {Shkolnik}, E.~L., {Ilyin}, I., {et~al.} 2019, \aj, 157, 69

\bibitem[{{Charbonneau} {et~al.}(2002){Charbonneau}, {Brown}, {Noyes}, \&
  {Gilliland}}]{2002ApJ...568..377C}
{Charbonneau}, D., {Brown}, T.~M., {Noyes}, R.~W., \& {Gilliland}, R.~L. 2002,
  \apj, 568, 377

\bibitem[{{Chen} {et~al.}(2020){Chen}, {Casasayas-Barris}, {Pall{\'e}}, {Yan},
  {Stangret}, {Cegla}, {Allart}, \& {Lovis}}]{2020A&A...635A.171C}
{Chen}, G., {Casasayas-Barris}, N., {Pall{\'e}}, E., {et~al.} 2020, \aap, 635,
  A171

\bibitem[{{Czesla} {et~al.}(2022){Czesla}, {Lamp{\'o}n}, {Sanz-Forcada},
  {Garc{\'\i}a Mu{\~n}oz}, {L{\'o}pez-Puertas}, {Nortmann}, {Yan}, {Nagel},
  {Yan}, {Schmitt}, {Aceituno}, {Amado}, {Caballero}, {Casasayas-Barris},
  {Henning}, {Khalafinejad}, {Molaverdikhani}, {Montes}, {Pall{\'e}},
  {Reiners}, {Schneider}, {Ribas}, {Quirrenbach}, {Zapatero Osorio}, \&
  {Zechmeister}}]{2022A&A...657A...6C}
{Czesla}, S., {Lamp{\'o}n}, M., {Sanz-Forcada}, J., {et~al.} 2022, \aap, 657,
  A6

\bibitem[{{Davis} \& {Wheatley}(2009)}]{2009MNRAS.396.1012D}
{Davis}, T.~A. \& {Wheatley}, P.~J. 2009, \mnras, 396, 1012

\bibitem[{{Debrecht} {et~al.}(2020){Debrecht}, {Carroll-Nellenback}, {Frank},
  {Blackman}, {Fossati}, {McCann}, \& {Murray-Clay}}]{2020MNRAS.493.1292D}
{Debrecht}, A., {Carroll-Nellenback}, J., {Frank}, A., {et~al.} 2020, \mnras,
  493, 1292

\bibitem[{{Deming} {et~al.}(2021){Deming}, {Stevenson}, \&
  {Ehrenreich}}]{2021exbi.book....7D}
{Deming}, D., {Stevenson}, K.~B., \& {Ehrenreich}, D. 2021, in ExoFrontiers;
  Big Questions in Exoplanetary Science, ed. N.~{Madhusudhan}, 7--1

\bibitem[{{Dos Santos} {et~al.}(2021){Dos Santos}, {Bourrier}, {Ehrenreich},
  {Sanz-Forcada}, {L{\'o}pez-Morales}, {Sing}, {Garc{\'\i}a Mu{\~n}oz},
  {Henry}, {Lavvas}, {Lecavelier des Etangs}, {Mikal-Evans}, {Vidal-Madjar}, \&
  {Wakeford}}]{DosSantos2021_WASP-29}
{Dos Santos}, L.~A., {Bourrier}, V., {Ehrenreich}, D., {et~al.} 2021, \aap,
  649, A40

\bibitem[{{Dos Santos} {et~al.}(2020{\natexlab{a}}){Dos Santos}, {Ehrenreich},
  {Bourrier}, {Allart}, {King}, {Lendl}, {Lovis}, {Margheim}, {Mel{\'e}ndez},
  {Seidel}, \& {Sousa}}]{DosSantos20}
{Dos Santos}, L.~A., {Ehrenreich}, D., {Bourrier}, V., {et~al.}
  2020{\natexlab{a}}, \aap, 640, A29

\bibitem[{{Dos Santos} {et~al.}(2020{\natexlab{b}}){Dos Santos}, {Ehrenreich},
  {Bourrier}, {Astudillo-Defru}, {Bonfils}, {Forget}, {Lovis}, {Pepe}, \&
  {Udry}}]{2020A&A...634L...4D}
{Dos Santos}, L.~A., {Ehrenreich}, D., {Bourrier}, V., {et~al.}
  2020{\natexlab{b}}, \aap, 634, L4

\bibitem[{{Dos Santos} {et~al.}(2019){Dos Santos}, {Ehrenreich}, {Bourrier},
  {Lecavelier des Etangs}, {L{\'o}pez-Morales}, {Sing}, {Ballester},
  {Ben-Jaffel}, {Buchhave}, {Garc{\'\i}a Mu{\~n}oz}, {Henry}, {Kataria},
  {Lavie}, {Lavvas}, {Lewis}, {Mikal-Evans}, {Sanz-Forcada}, \&
  {Wakeford}}]{DosSantos19}
{Dos Santos}, L.~A., {Ehrenreich}, D., {Bourrier}, V., {et~al.} 2019, \aap,
  629, A47

\bibitem[{{Dos Santos} {et~al.}(2022){Dos Santos}, {Vidotto}, {Vissapragada},
  {Alam}, {Allart}, {Bourrier}, {Kirk}, {Seidel}, \&
  {Ehrenreich}}]{2022A&A...659A..62D}
{Dos Santos}, L.~A., {Vidotto}, A.~A., {Vissapragada}, S., {et~al.} 2022, \aap,
  659, A62

\bibitem[{{Ehrenreich} {et~al.}(2012){Ehrenreich}, {Bourrier}, {Bonfils},
  {Lecavelier des Etangs}, {H{\'e}brard}, {Sing}, {Wheatley}, {Vidal-Madjar},
  {Delfosse}, {Udry}, {Forveille}, \& {Moutou}}]{2012A&A...547A..18E}
{Ehrenreich}, D., {Bourrier}, V., {Bonfils}, X., {et~al.} 2012, \aap, 547, A18

\bibitem[{{Ehrenreich} {et~al.}(2015){Ehrenreich}, {Bourrier}, {Wheatley},
  {Lecavelier des Etangs}, {H{\'e}brard}, {Udry}, {Bonfils}, {Delfosse},
  {D{\'e}sert}, {Sing}, \& {Vidal-Madjar}}]{2015Natur.522..459E}
{Ehrenreich}, D., {Bourrier}, V., {Wheatley}, P.~J., {et~al.} 2015, \nat, 522,
  459

\bibitem[{{Ehrenreich} {et~al.}(2011){Ehrenreich}, {Lecavelier Des Etangs}, \&
  {Delfosse}}]{2011A&A...529A..80E}
{Ehrenreich}, D., {Lecavelier Des Etangs}, A., \& {Delfosse}, X. 2011, \aap,
  529, A80

\bibitem[{{Ehrenreich} {et~al.}(2008){Ehrenreich}, {Lecavelier Des Etangs},
  {H{\'e}brard}, {D{\'e}sert}, {Vidal-Madjar}, {McConnell}, {Parkinson},
  {Ballester}, \& {Ferlet}}]{2008A&A...483..933E}
{Ehrenreich}, D., {Lecavelier Des Etangs}, A., {H{\'e}brard}, G., {et~al.}
  2008, \aap, 483, 933

\bibitem[{{Erkaev} {et~al.}(2007){Erkaev}, {Kulikov}, {Lammer}, {Selsis},
  {Langmayr}, {Jaritz}, \& {Biernat}}]{2007A&A...472..329E}
{Erkaev}, N.~V., {Kulikov}, Y.~N., {Lammer}, H., {et~al.} 2007, \aap, 472, 329

\bibitem[{{Erkaev} {et~al.}(2016){Erkaev}, {Lammer}, {Odert}, {Kislyakova},
  {Johnstone}, {G{\"u}del}, \& {Khodachenko}}]{2016MNRAS.460.1300E}
{Erkaev}, N.~V., {Lammer}, H., {Odert}, P., {et~al.} 2016, \mnras, 460, 1300

\bibitem[{{Fossati} {et~al.}(2015){Fossati}, {France}, {Koskinen}, {Juvan},
  {Haswell}, \& {Lendl}}]{2015ApJ...815..118F}
{Fossati}, L., {France}, K., {Koskinen}, T., {et~al.} 2015, \apj, 815, 118

\bibitem[{{Fossati} {et~al.}(2022){Fossati}, {Guilluy}, {Shaikhislamov},
  {Carleo}, {Borsa}, {Bonomo}, {Giacobbe}, {Rainer}, {Cecchi-Pestellini},
  {Khodachenko}, {Efimov}, {Rumenskikh}, {Miroshnichenko}, {Berezutsky},
  {Nascimbeni}, {Brogi}, {Lanza}, {Mancini}, {Affer}, {Benatti}, {Biazzo},
  {Bignamini}, {Carosati}, {Claudi}, {Cosentino}, {Covino}, {Desidera},
  {Fiorenzano}, {Harutyunyan}, {Maggio}, {Malavolta}, {Maldonado}, {Micela},
  {Molinari}, {Pagano}, {Pedani}, {Piotto}, {Poretti}, {Scandariato},
  {Sozzetti}, \& {Stoev}}]{Fossati22}
{Fossati}, L., {Guilluy}, G., {Shaikhislamov}, I.~F., {et~al.} 2022, \aap, 658,
  A136

\bibitem[{{Fossati} {et~al.}(2010){Fossati}, {Haswell}, {Froning}, {Hebb},
  {Holmes}, {Kolb}, {Helling}, {Carter}, {Wheatley}, {Collier Cameron},
  {Loeillet}, {Pollacco}, {Street}, {Stempels}, {Simpson}, {Udry}, {Joshi},
  {West}, {Skillen}, \& {Wilson}}]{2010ApJ...714L.222F}
{Fossati}, L., {Haswell}, C.~A., {Froning}, C.~S., {et~al.} 2010, \apjl, 714,
  L222

\bibitem[{{Fossati} {et~al.}(2018){Fossati}, {Koskinen}, {France}, {Cubillos},
  {Haswell}, {Lanza}, \& {Pillitteri}}]{2018AJ....155..113F}
{Fossati}, L., {Koskinen}, T., {France}, K., {et~al.} 2018, \aj, 155, 113

\bibitem[{{France} {et~al.}(2020){France}, {Duvvuri}, {Egan}, {Koskinen},
  {Wilson}, {Youngblood}, {Froning}, {Brown}, {Alvarado-G{\'o}mez},
  {Berta-Thompson}, {Drake}, {Garraffo}, {Kaltenegger}, {Kowalski}, {Linsky},
  {Loyd}, {Mauas}, {Miguel}, {Pineda}, {Rugheimer}, {Schneider}, {Tian}, \&
  {Vieytes}}]{2020AJ....160..237F}
{France}, K., {Duvvuri}, G., {Egan}, H., {et~al.} 2020, \aj, 160, 237

\bibitem[{{Gaidos} {et~al.}(2020){Gaidos}, {Hirano}, {Mann}, {Owens}, {Berger},
  {France}, {Vanderburg}, {Harakawa}, {Hodapp}, {Ishizuka}, {Jacobson},
  {Konishi}, {Kotani}, {Kudo}, {Kurokawa}, {Kuzuhara}, {Nishikawa}, {Omiya},
  {Serizawa}, {Tamura}, \& {Ueda}}]{2020MNRAS.495..650G}
{Gaidos}, E., {Hirano}, T., {Mann}, A.~W., {et~al.} 2020, \mnras, 495, 650

\bibitem[{{Gaidos} {et~al.}(2021){Gaidos}, {Hirano}, {Omiya}, {Kuzuhara},
  {Kotani}, {Tamura}, {Harakawa}, \& {Kudo}}]{2021RNAAS...5..238G}
{Gaidos}, E., {Hirano}, T., {Omiya}, M., {et~al.} 2021, Research Notes of the
  American Astronomical Society, 5, 238

\bibitem[{{Garc{\'\i}a Mu{\~n}oz} {et~al.}(2021){Garc{\'\i}a Mu{\~n}oz},
  {Fossati}, {Youngblood}, {Nettelmann}, {Gandolfi}, {Cabrera}, \&
  {Rauer}}]{2021ApJ...907L..36G}
{Garc{\'\i}a Mu{\~n}oz}, A., {Fossati}, L., {Youngblood}, A., {et~al.} 2021,
  \apjl, 907, L36

\bibitem[{{Garc{\'\i}a Mu{\~n}oz} {et~al.}(2020){Garc{\'\i}a Mu{\~n}oz},
  {Youngblood}, {Fossati}, {Gandolfi}, {Cabrera}, \&
  {Rauer}}]{2020ApJ...888L..21G}
{Garc{\'\i}a Mu{\~n}oz}, A., {Youngblood}, A., {Fossati}, L., {et~al.} 2020,
  \apjl, 888, L21

\bibitem[{{Ginzburg} {et~al.}(2018){Ginzburg}, {Schlichting}, \&
  {Sari}}]{2018MNRAS.476..759G}
{Ginzburg}, S., {Schlichting}, H.~E., \& {Sari}, R. 2018, \mnras, 476, 759

\bibitem[{{Gross}(1972)}]{1972JAtS...29..214G}
{Gross}, S.~H. 1972, Journal of Atmospheric Sciences, 29, 214

\bibitem[{{Guillot} {et~al.}(1996){Guillot}, {Burrows}, {Hubbard}, {Lunine}, \&
  {Saumon}}]{1996ApJ...459L..35G}
{Guillot}, T., {Burrows}, A., {Hubbard}, W.~B., {Lunine}, J.~I., \& {Saumon},
  D. 1996, \apjl, 459, L35

\bibitem[{{Guilluy} {et~al.}(2020){Guilluy}, {Andretta}, {Borsa}, {Giacobbe},
  {Sozzetti}, {Covino}, {Bourrier}, {Fossati}, {Bonomo}, {Esposito},
  {Giampapa}, {Harutyunyan}, {Rainer}, {Brogi}, {Bruno}, {Claudi}, {Frustagli},
  {Lanza}, {Mancini}, {Pino}, {Poretti}, {Scandariato}, {Affer}, {Baffa},
  {Baruffolo}, {Benatti}, {Biazzo}, {Bignamini}, {Boschin}, {Carleo},
  {Cecconi}, {Cosentino}, {Damasso}, {Desidera}, {Falcini}, {Martinez
  Fiorenzano}, {Ghedina}, {Gonz{\'a}lez-{\'A}lvarez}, {Guerra}, {Hernandez},
  {Leto}, {Maggio}, {Malavolta}, {Maldonado}, {Micela}, {Molinari},
  {Nascimbeni}, {Pagano}, {Pedani}, {Piotto}, \&
  {Reiners}}]{2020A&A...639A..49G}
{Guilluy}, G., {Andretta}, V., {Borsa}, F., {et~al.} 2020, \aap, 639, A49

\bibitem[{{Gupta} \& {Schlichting}(2019)}]{2019MNRAS.487...24G}
{Gupta}, A. \& {Schlichting}, H.~E. 2019, \mnras, 487, 24

\bibitem[{{Gupta} \& {Schlichting}(2020)}]{2020MNRAS.493..792G}
{Gupta}, A. \& {Schlichting}, H.~E. 2020, \mnras, 493, 792

\bibitem[{{Hirano} {et~al.}(2020){Hirano}, {Krishnamurthy}, {Gaidos},
  {Flewelling}, {Mann}, {Narita}, {Plavchan}, {Kotani}, {Tamura}, {Harakawa},
  {Hodapp}, {Ishizuka}, {Jacobson}, {Konishi}, {Kudo}, {Kurokawa}, {Kuzuhara},
  {Nishikawa}, {Omiya}, {Serizawa}, {Ueda}, \& {Vievard}}]{2020ApJ...899L..13H}
{Hirano}, T., {Krishnamurthy}, V., {Gaidos}, E., {et~al.} 2020, \apjl, 899, L13

\bibitem[{{Holmstr{\"o}m} {et~al.}(2008){Holmstr{\"o}m}, {Ekenb{\"a}ck},
  {Selsis}, {Penz}, {Lammer}, \& {Wurz}}]{2008Natur.451..970H}
{Holmstr{\"o}m}, M., {Ekenb{\"a}ck}, A., {Selsis}, F., {et~al.} 2008, \nat,
  451, 970

\bibitem[{{Jensen} {et~al.}(2018){Jensen}, {Cauley}, {Redfield}, {Cochran}, \&
  {Endl}}]{2018AJ....156..154J}
{Jensen}, A.~G., {Cauley}, P.~W., {Redfield}, S., {Cochran}, W.~D., \& {Endl},
  M. 2018, \aj, 156, 154

\bibitem[{{Jensen} {et~al.}(2012){Jensen}, {Redfield}, {Endl}, {Cochran},
  {Koesterke}, \& {Barman}}]{2012ApJ...751...86J}
{Jensen}, A.~G., {Redfield}, S., {Endl}, M., {et~al.} 2012, \apj, 751, 86

\bibitem[{{Kasper} {et~al.}(2020){Kasper}, {Bean}, {Oklop{\v{c}}i{\'c}},
  {Malsky}, {Kempton}, {D{\'e}sert}, {Rogers}, \&
  {Mansfield}}]{2020AJ....160..258K}
{Kasper}, D., {Bean}, J.~L., {Oklop{\v{c}}i{\'c}}, A., {et~al.} 2020, \aj, 160,
  258

\bibitem[{{Khodachenko} {et~al.}(2019){Khodachenko}, {Shaikhislamov}, {Lammer},
  {Berezutsky}, {Miroshnichenko}, {Rumenskikh}, {Kislyakova}, \&
  {Dwivedi}}]{2019ApJ...885...67K}
{Khodachenko}, M.~L., {Shaikhislamov}, I.~F., {Lammer}, H., {et~al.} 2019,
  \apj, 885, 67

\bibitem[{{King} {et~al.}(2021){King}, {Corrales}, {Wheatley}, {Lavvas},
  {Steinrueck}, {Bourrier}, {Ehrenreich}, {Lecavelier des Etangs}, \&
  {Louden}}]{2021MNRAS.506.2453K}
{King}, G.~W., {Corrales}, L., {Wheatley}, P.~J., {et~al.} 2021, \mnras, 506,
  2453

\bibitem[{{King} \& {Wheatley}(2021)}]{2021MNRAS.501L..28K}
{King}, G.~W. \& {Wheatley}, P.~J. 2021, \mnras, 501, L28

\bibitem[{{Kirk} {et~al.}(2020){Kirk}, {Alam}, {L{\'o}pez-Morales}, \&
  {Zeng}}]{2020AJ....159..115K}
{Kirk}, J., {Alam}, M.~K., {L{\'o}pez-Morales}, M., \& {Zeng}, L. 2020, \aj,
  159, 115

\bibitem[{{Kirk} {et~al.}(2022){Kirk}, {Dos Santos}, {L{\'o}pez-Morales},
  {Alam}, {Oklop{\v{c}}i{\'c}}, {MacLeod}, {Zeng}, \&
  {Zhou}}]{2022AJ....164...24K}
{Kirk}, J., {Dos Santos}, L.~A., {L{\'o}pez-Morales}, M., {et~al.} 2022, \aj,
  164, 24

\bibitem[{{Kislyakova} {et~al.}(2019){Kislyakova}, {Holmstr{\"o}m}, {Odert},
  {Lammer}, {Erkaev}, {Khodachenko}, {Shaikhislamov}, {Dorfi}, \&
  {G{\"u}del}}]{2019A&A...623A.131K}
{Kislyakova}, K.~G., {Holmstr{\"o}m}, M., {Odert}, P., {et~al.} 2019, \aap,
  623, A131

\bibitem[{{Koskinen} {et~al.}(2007){Koskinen}, {Aylward}, \&
  {Miller}}]{2007Natur.450..845K}
{Koskinen}, T.~T., {Aylward}, A.~D., \& {Miller}, S. 2007, \nat, 450, 845

\bibitem[{{Kreidberg} \& {Oklop{\v{c}}i{\'c}}(2018)}]{2018RNAAS...2...44K}
{Kreidberg}, L. \& {Oklop{\v{c}}i{\'c}}, A. 2018, Research Notes of the
  American Astronomical Society, 2, 44

\bibitem[{{Krishnamurthy} {et~al.}(2021){Krishnamurthy}, {Hirano},
  {Stef{\'a}nsson}, {Ninan}, {Mahadevan}, {Gaidos}, {Kopparapu}, {Sato},
  {Hori}, {Bender}, {Ca{\~n}as}, {Diddams}, {Halverson}, {Harakawa}, {Hawley},
  {Hearty}, {Hebb}, {Hodapp}, {Jacobson}, {Kanodia}, {Konishi}, {Kotani},
  {Kowalski}, {Kudo}, {Kurokawa}, {Kuzuhara}, {Lin}, {Maney}, {Metcalf},
  {Morris}, {Nishikawa}, {Omiya}, {Robertson}, {Roy}, {Schwab}, {Serizawa},
  {Tamura}, {Ueda}, {Vievard}, \& {Wisniewski}}]{2021AJ....162...82K}
{Krishnamurthy}, V., {Hirano}, T., {Stef{\'a}nsson}, G., {et~al.} 2021, \aj,
  162, 82

\bibitem[{{Kulow} {et~al.}(2014){Kulow}, {France}, {Linsky}, \&
  {Loyd}}]{2014ApJ...786..132K}
{Kulow}, J.~R., {France}, K., {Linsky}, J., \& {Loyd}, R.~O.~P. 2014, \apj,
  786, 132

\bibitem[{{Lammer} {et~al.}(2003){Lammer}, {Selsis}, {Ribas}, {Guinan},
  {Bauer}, \& {Weiss}}]{2003ApJ...598L.121L}
{Lammer}, H., {Selsis}, F., {Ribas}, I., {et~al.} 2003, \apjl, 598, L121

\bibitem[{{Lamp{\'o}n} {et~al.}(2021){Lamp{\'o}n}, {L{\'o}pez-Puertas},
  {Czesla}, {S{\'a}nchez-L{\'o}pez}, {Lara}, {Salz}, {Sanz-Forcada},
  {Molaverdikhani}, {Quirrenbach}, {Pall{\'e}}, {Caballero}, {Henning},
  {Nortmann}, {Amado}, {Montes}, {Reiners}, \& {Ribas}}]{2021A&A...648L...7L}
{Lamp{\'o}n}, M., {L{\'o}pez-Puertas}, M., {Czesla}, S., {et~al.} 2021, \aap,
  648, L7

\bibitem[{{Lamp{\'o}n} {et~al.}(2020){Lamp{\'o}n}, {L{\'o}pez-Puertas}, {Lara},
  {S{\'a}nchez-L{\'o}pez}, {Salz}, {Czesla}, {Sanz-Forcada}, {Molaverdikhani},
  {Alonso-Floriano}, {Nortmann}, {Caballero}, {Bauer}, {Pall{\'e}}, {Montes},
  {Quirrenbach}, {Nagel}, {Ribas}, {Reiners}, \& {Amado}}]{2020A&A...636A..13L}
{Lamp{\'o}n}, M., {L{\'o}pez-Puertas}, M., {Lara}, L.~M., {et~al.} 2020, \aap,
  636, A13

\bibitem[{{Lavie} {et~al.}(2017){Lavie}, {Ehrenreich}, {Bourrier}, {Lecavelier
  des Etangs}, {Vidal-Madjar}, {Delfosse}, {Gracia Berna}, {Heng}, {Thomas},
  {Udry}, \& {Wheatley}}]{2017A&A...605L...7L}
{Lavie}, B., {Ehrenreich}, D., {Bourrier}, V., {et~al.} 2017, \aap, 605, L7

\bibitem[{{Lecavelier Des Etangs}(2007)}]{2007A&A...461.1185L}
{Lecavelier Des Etangs}, A. 2007, \aap, 461, 1185

\bibitem[{{Lecavelier des Etangs} {et~al.}(2012){Lecavelier des Etangs},
  {Bourrier}, {Wheatley}, {Dupuy}, {Ehrenreich}, {Vidal-Madjar}, {H{\'e}brard},
  {Ballester}, {D{\'e}sert}, {Ferlet}, \& {Sing}}]{2012A&A...543L...4L}
{Lecavelier des Etangs}, A., {Bourrier}, V., {Wheatley}, P.~J., {et~al.} 2012,
  \aap, 543, L4

\bibitem[{{Lecavelier des Etangs} {et~al.}(2010){Lecavelier des Etangs},
  {Ehrenreich}, {Vidal-Madjar}, {Ballester}, {D{\'e}sert}, {Ferlet},
  {H{\'e}brard}, {Sing}, {Tchakoumegni}, \& {Udry}}]{2010A&A...514A..72L}
{Lecavelier des Etangs}, A., {Ehrenreich}, D., {Vidal-Madjar}, A., {et~al.}
  2010, \aap, 514, A72

\bibitem[{{Lecavelier Des Etangs} {et~al.}(2008){Lecavelier Des Etangs},
  {Vidal-Madjar}, \& {Desert}}]{2008Natur.456E...1L}
{Lecavelier Des Etangs}, A., {Vidal-Madjar}, A., \& {Desert}, J.~M. 2008, \nat,
  456, E1

\bibitem[{{Lecavelier des Etangs} {et~al.}(2004){Lecavelier des Etangs},
  {Vidal-Madjar}, {McConnell}, \& {H{\'e}brard}}]{2004A&A...418L...1L}
{Lecavelier des Etangs}, A., {Vidal-Madjar}, A., {McConnell}, J.~C., \&
  {H{\'e}brard}, G. 2004, \aap, 418, L1

\bibitem[{{Linsky} {et~al.}(2010){Linsky}, {Yang}, {France}, {Froning},
  {Green}, {Stocke}, \& {Osterman}}]{2010ApJ...717.1291L}
{Linsky}, J.~L., {Yang}, H., {France}, K., {et~al.} 2010, \apj, 717, 1291

\bibitem[{{Linssen} {et~al.}(2022){Linssen}, {Oklop{\v{c}}i{\'c}}, \&
  {MacLeod}}]{2022arXiv220903677L}
{Linssen}, D., {Oklop{\v{c}}i{\'c}}, A., \& {MacLeod}, M. 2022, arXiv e-prints,
  arXiv:2209.03677

\bibitem[{{Loyd} {et~al.}(2017){Loyd}, {Koskinen}, {France}, {Schneider}, \&
  {Redfield}}]{Loyd17}
{Loyd}, R.~O.~P., {Koskinen}, T.~T., {France}, K., {Schneider}, C., \&
  {Redfield}, S. 2017, \apjl, 834, L17

\bibitem[{{Mallonn} \& {Strassmeier}(2016)}]{Mallonn16}
{Mallonn}, M. \& {Strassmeier}, K.~G. 2016, \aap, 590, A100

\bibitem[{{Mansfield} {et~al.}(2018){Mansfield}, {Bean}, {Oklop{\v{c}}i{\'c}},
  {Kreidberg}, {D{\'e}sert}, {Kempton}, {Line}, {Fortney}, {Henry}, {Mallonn},
  {Stevenson}, {Dragomir}, {Allart}, \& {Bourrier}}]{2018ApJ...868L..34M}
{Mansfield}, M., {Bean}, J.~L., {Oklop{\v{c}}i{\'c}}, A., {et~al.} 2018, \apjl,
  868, L34

\bibitem[{{Mayor} \& {Queloz}(1995)}]{1995Natur.378..355M}
{Mayor}, M. \& {Queloz}, D. 1995, \nat, 378, 355

\bibitem[{{Mazeh} {et~al.}(2016){Mazeh}, {Holczer}, \&
  {Faigler}}]{2016A&A...589A..75M}
{Mazeh}, T., {Holczer}, T., \& {Faigler}, S. 2016, \aap, 589, A75

\bibitem[{{Moutou} {et~al.}(2003){Moutou}, {Coustenis}, {Schneider}, {Queloz},
  \& {Mayor}}]{Moutou03}
{Moutou}, C., {Coustenis}, A., {Schneider}, J., {Queloz}, D., \& {Mayor}, M.
  2003, \aap, 405, 341

\bibitem[{{Murray-Clay} {et~al.}(2009){Murray-Clay}, {Chiang}, \&
  {Murray}}]{2009ApJ...693...23M}
{Murray-Clay}, R.~A., {Chiang}, E.~I., \& {Murray}, N. 2009, \apj, 693, 23

\bibitem[{{National Academies of Sciences, Engineering, and
  Medicine}(2021)}]{NAP26141}
{National Academies of Sciences, Engineering, and Medicine}. 2021, Pathways to
  Discovery in Astronomy and Astrophysics for the 2020s (Washington, DC: The
  National Academies Press)

\bibitem[{{Ninan} {et~al.}(2020){Ninan}, {Stefansson}, {Mahadevan}, {Bender},
  {Robertson}, {Ramsey}, {Terrien}, {Wright}, {Diddams}, {Kanodia}, {Cochran},
  {Endl}, {Ford}, {Fredrick}, {Halverson}, {Hearty}, {Jennings}, {Kaplan},
  {Lubar}, {Metcalf}, {Monson}, {Nitroy}, {Roy}, \&
  {Schwab}}]{2020ApJ...894...97N}
{Ninan}, J.~P., {Stefansson}, G., {Mahadevan}, S., {et~al.} 2020, \apj, 894, 97

\bibitem[{{Nortmann} {et~al.}(2018){Nortmann}, {Pall{\'e}}, {Salz},
  {Sanz-Forcada}, {Nagel}, {Alonso-Floriano}, {Czesla}, {Yan}, {Chen},
  {Snellen}, {Zechmeister}, {Schmitt}, {L{\'o}pez-Puertas}, {Casasayas-Barris},
  {Bauer}, {Amado}, {Caballero}, {Dreizler}, {Henning}, {Lamp{\'o}n}, {Montes},
  {Molaverdikhani}, {Quirrenbach}, {Reiners}, {Ribas}, {S{\'a}nchez-L{\'o}pez},
  {Schneider}, \& {Zapatero Osorio}}]{2018Sci...362.1388N}
{Nortmann}, L., {Pall{\'e}}, E., {Salz}, M., {et~al.} 2018, Science, 362, 1388

\bibitem[{{Oklop{\v{c}}i{\'c}}(2019)}]{2019ApJ...881..133O}
{Oklop{\v{c}}i{\'c}}, A. 2019, \apj, 881, 133

\bibitem[{{Oklop{\v{c}}i{\'c}} \& {Hirata}(2018)}]{2018ApJ...855L..11O}
{Oklop{\v{c}}i{\'c}}, A. \& {Hirata}, C.~M. 2018, \apjl, 855, L11

\bibitem[{{{\"O}pik}(1963)}]{1963GeoJ....7..490O}
{{\"O}pik}, E.~J. 1963, Geophysical Journal, 7, 490

\bibitem[{{Orell-Miquel} {et~al.}(2022){Orell-Miquel}, {Murgas}, {Pall{\'e}},
  {Lamp{\'o}n}, {L{\'o}pez-Puertas}, {Sanz-Forcada}, {Nagel}, {Kaminski},
  {Casasayas-Barris}, {Nortmann}, {Luque}, {Molaverdikhani}, {Sedaghati},
  {Caballero}, {Amado}, {Bergond}, {Czesla}, {Hatzes}, {Henning},
  {Khalafinejad}, {Montes}, {Morello}, {Quirrenbach}, {Reiners}, {Ribas},
  {S{\'a}nchez-L{\'o}pez}, {Schweitzer}, {Stangret}, {Yan}, \& {Zapatero
  Osorio}}]{2022A&A...659A..55O}
{Orell-Miquel}, J., {Murgas}, F., {Pall{\'e}}, E., {et~al.} 2022, \aap, 659,
  A55

\bibitem[{{Owen} \& {Alvarez}(2016)}]{2016ApJ...816...34O}
{Owen}, J.~E. \& {Alvarez}, M.~A. 2016, \apj, 816, 34

\bibitem[{{Owen} \& {Lai}(2018)}]{2018MNRAS.479.5012O}
{Owen}, J.~E. \& {Lai}, D. 2018, \mnras, 479, 5012

\bibitem[{{Owen} {et~al.}(2021){Owen}, {Murray-Clay}, {Schreyer},
  {Schlichting}, {Ardila}, {Gupta}, {Loyd}, {Shkolnik}, {Sing}, \&
  {Swain}}]{2021arXiv211106094O}
{Owen}, J.~E., {Murray-Clay}, R.~A., {Schreyer}, E., {et~al.} 2021, arXiv
  e-prints, arXiv:2111.06094

\bibitem[{{Owen} \& {Wu}(2013)}]{2013ApJ...775..105O}
{Owen}, J.~E. \& {Wu}, Y. 2013, \apj, 775, 105

\bibitem[{{Palle} {et~al.}(2020){Palle}, {Nortmann}, {Casasayas-Barris},
  {Lamp{\'o}n}, {L{\'o}pez-Puertas}, {Caballero}, {Sanz-Forcada}, {Lara},
  {Nagel}, {Yan}, {Alonso-Floriano}, {Amado}, {Chen}, {Cifuentes},
  {Cort{\'e}s-Contreras}, {Czesla}, {Molaverdikhani}, {Montes}, {Passegger},
  {Quirrenbach}, {Reiners}, {Ribas}, {S{\'a}nchez-L{\'o}pez}, {Schweitzer},
  {Stangret}, {Zapatero Osorio}, \& {Zechmeister}}]{2020A&A...638A..61P}
{Palle}, E., {Nortmann}, L., {Casasayas-Barris}, N., {et~al.} 2020, \aap, 638,
  A61

\bibitem[{{Paragas} {et~al.}(2021){Paragas}, {Vissapragada}, {Knutson},
  {Oklop{\v{c}}i{\'c}}, {Chachan}, {Greklek-McKeon}, {Dai}, {Tinyanont}, \&
  {Vasisht}}]{2021ApJ...909L..10P}
{Paragas}, K., {Vissapragada}, S., {Knutson}, H.~A., {et~al.} 2021, \apjl, 909,
  L10

\bibitem[{{Parker}(1958)}]{1958ApJ...128..664P}
{Parker}, E.~N. 1958, \apj, 128, 664

\bibitem[{{Pillitteri} {et~al.}(2022){Pillitteri}, {Micela}, {Maggio},
  {Sciortino}, \& {Lopez-Santiago}}]{2022A&A...660A..75P}
{Pillitteri}, I., {Micela}, G., {Maggio}, A., {Sciortino}, S., \&
  {Lopez-Santiago}, J. 2022, \aap, 660, A75

\bibitem[{{Poppenhaeger}(2022)}]{2022MNRAS.512.1751P}
{Poppenhaeger}, K. 2022, \mnras, 512, 1751

\bibitem[{{Rackham} {et~al.}(2019){Rackham}, {Apai}, \&
  {Giampapa}}]{2019AJ....157...96R}
{Rackham}, B.~V., {Apai}, D., \& {Giampapa}, M.~S. 2019, \aj, 157, 96

\bibitem[{{Rackham} {et~al.}(2022){Rackham}, {Espinoza}, {Berdyugina},
  {Korhonen}, {MacDonald}, {Montet}, {Morris}, {Oshagh}, {Shapiro}, {Unruh},
  {Quintana}, {Zellem}, {Apai}, {Barclay}, {Barstow}, {Bruno}, {Carone},
  {Casewell}, {Cegla}, {Criscuoli}, {Fischer}, {Fournier}, {Giampapa}, {Giles},
  {Iyer}, {Kopp}, {Kostogryz}, {Krivova}, {Mallonn}, {McGruder},
  {Molaverdikhani}, {Newton}, {Panja}, {Peacock}, {Reardon}, {Roettenbacher},
  {Scandariato}, {Solanki}, {Stassun}, {Steiner}, {Stevenson}, {Tregloan-Reed},
  {Valio}, {Wedemeyer}, {Welbanks}, {Yu}, {Alam}, {Davenport}, {Deming},
  {Dong}, {Ducrot}, {Fisher}, {Gilbert}, {Kostov}, {L{\'o}pez-Morales}, {Line},
  {Mo{\v{c}}nik}, {Mullally}, {Paudel}, {Ribas}, \&
  {Valenti}}]{2022arXiv220109905R}
{Rackham}, B.~V., {Espinoza}, N., {Berdyugina}, S.~V., {et~al.} 2022, arXiv
  e-prints, arXiv:2201.09905

\bibitem[{{Ridden-Harper} {et~al.}(2016){Ridden-Harper}, {Snellen}, {Keller},
  {de Kok}, {Di Gloria}, {Hoeijmakers}, {Brogi}, {Fridlund}, {Vermeersen}, \&
  {van Westrenen}}]{Ridden16}
{Ridden-Harper}, A.~R., {Snellen}, I.~A.~G., {Keller}, C.~U., {et~al.} 2016,
  \aap, 593, A129

\bibitem[{{Rockcliffe} {et~al.}(2021){Rockcliffe}, {Newton}, {Youngblood},
  {Bourrier}, {Mann}, {Berta-Thompson}, {Ag{\"u}eros}, {N{\'u}{\~n}ez}, \&
  {Charbonneau}}]{2021AJ....162..116R}
{Rockcliffe}, K.~E., {Newton}, E.~R., {Youngblood}, A., {et~al.} 2021, \aj,
  162, 116

\bibitem[{{Salz} {et~al.}(2018){Salz}, {Czesla}, {Schneider}, {Nagel},
  {Schmitt}, {Nortmann}, {Alonso-Floriano}, {L{\'o}pez-Puertas}, {Lamp{\'o}n},
  {Bauer}, {Snellen}, {Pall{\'e}}, {Caballero}, {Yan}, {Chen}, {Sanz-Forcada},
  {Amado}, {Quirrenbach}, {Ribas}, {Reiners}, {B{\'e}jar}, {Casasayas-Barris},
  {Cort{\'e}s-Contreras}, {Dreizler}, {Guenther}, {Henning}, {Jeffers},
  {Kaminski}, {K{\"u}rster}, {Lafarga}, {Lara}, {Molaverdikhani}, {Montes},
  {Morales}, {S{\'a}nchez-L{\'o}pez}, {Seifert}, {Zapatero Osorio}, \&
  {Zechmeister}}]{2018A&A...620A..97S}
{Salz}, M., {Czesla}, S., {Schneider}, P.~C., {et~al.} 2018, \aap, 620, A97

\bibitem[{{Salz} {et~al.}(2016){Salz}, {Schneider}, {Czesla}, \&
  {Schmitt}}]{2016A&A...585L...2S}
{Salz}, M., {Schneider}, P.~C., {Czesla}, S., \& {Schmitt}, J.~H.~M.~M. 2016,
  \aap, 585, L2

\bibitem[{{S{\'a}nchez-L{\'o}pez} {et~al.}(2022){S{\'a}nchez-L{\'o}pez}, {Lin},
  {Snellen}, {Casasayas-Barris}, {Garc{\'\i}a Mu{\~n}oz}, {Lamp{\'o}n}, \&
  {L{\'o}pez-Puertas}}]{2022A&A...666L...1S}
{S{\'a}nchez-L{\'o}pez}, A., {Lin}, L., {Snellen}, I.~A.~G., {et~al.} 2022,
  \aap, 666, L1

\bibitem[{{Schilling}(1996)}]{1996Sci...273..429S}
{Schilling}, G. 1996, Science, 273, 429

\bibitem[{{Schlawin} {et~al.}(2010){Schlawin}, {Agol}, {Walkowicz}, {Covey}, \&
  {Lloyd}}]{2010ApJ...722L..75S}
{Schlawin}, E., {Agol}, E., {Walkowicz}, L.~M., {Covey}, K., \& {Lloyd}, J.~P.
  2010, \apjl, 722, L75

\bibitem[{{Seager} \& {Sasselov}(2000)}]{2000ApJ...537..916S}
{Seager}, S. \& {Sasselov}, D.~D. 2000, \apj, 537, 916

\bibitem[{{Sing} {et~al.}(2016){Sing}, {Fortney}, {Nikolov}, {Wakeford},
  {Kataria}, {Evans}, {Aigrain}, {Ballester}, {Burrows}, {Deming},
  {D{\'e}sert}, {Gibson}, {Henry}, {Huitson}, {Knutson}, {Lecavelier Des
  Etangs}, {Pont}, {Showman}, {Vidal-Madjar}, {Williamson}, \&
  {Wilson}}]{2016Natur.529...59S}
{Sing}, D.~K., {Fortney}, J.~J., {Nikolov}, N., {et~al.} 2016, \nat, 529, 59

\bibitem[{{Sing} {et~al.}(2019){Sing}, {Lavvas}, {Ballester}, {Lecavelier des
  Etangs}, {Marley}, {Nikolov}, {Ben-Jaffel}, {Bourrier}, {Buchhave}, {Deming},
  {Ehrenreich}, {Mikal-Evans}, {Kataria}, {Lewis}, {L{\'o}pez-Morales},
  {Garc{\'\i}a Mu{\~n}oz}, {Henry}, {Sanz-Forcada}, {Spake}, {Wakeford}, \&
  {PanCET Collaboration}}]{2019AJ....158...91S}
{Sing}, D.~K., {Lavvas}, P., {Ballester}, G.~E., {et~al.} 2019, \aj, 158, 91

\bibitem[{{Spake} {et~al.}(2021){Spake}, {Oklop{\v{c}}i{\'c}}, \&
  {Hillenbrand}}]{2021AJ....162..284S}
{Spake}, J.~J., {Oklop{\v{c}}i{\'c}}, A., \& {Hillenbrand}, L.~A. 2021, \aj,
  162, 284

\bibitem[{{Spake} {et~al.}(2022){Spake}, {Oklop{\v{c}}i{\'c}}, {Hillenbrand},
  {Knutson}, {Kasper}, {Dai}, {Orell-Miquel}, {Vissapragada}, {Zhang}, \&
  {Bean}}]{2022arXiv220903502S}
{Spake}, J.~J., {Oklop{\v{c}}i{\'c}}, A., {Hillenbrand}, L.~A., {et~al.} 2022,
  arXiv e-prints, arXiv:2209.03502

\bibitem[{{Spake} {et~al.}(2018){Spake}, {Sing}, {Evans}, {Oklop{\v{c}}i{\'c}},
  {}, {Bourrier}, {Kreidberg}, {Rackham}, {Irwin}, {Ehrenreich}, {Wyttenbach},
  {Wakeford}, {Zhou}, {Chubb}, {Nikolov}, {Goyal}, {Henry}, {Williamson},
  {Blumenthal}, {Anderson}, {Hellier}, {Charbonneau}, {Udry}, \&
  {Madhusudhan}}]{2018Natur.557...68S}
{Spake}, J.~J., {Sing}, D.~K., {Evans}, T.~M., {et~al.} 2018, \nat, 557, 68

\bibitem[{{Szab{\'o}} \& {Kiss}(2011)}]{2011ApJ...727L..44S}
{Szab{\'o}}, G.~M. \& {Kiss}, L.~L. 2011, \apj, 727, L44

\bibitem[{{Tripathi} {et~al.}(2015){Tripathi}, {Kratter}, {Murray-Clay}, \&
  {Krumholz}}]{2015ApJ...808..173T}
{Tripathi}, A., {Kratter}, K.~M., {Murray-Clay}, R.~A., \& {Krumholz}, M.~R.
  2015, \apj, 808, 173

\bibitem[{{Vidal-Madjar} {et~al.}(2004){Vidal-Madjar}, {D{\'e}sert},
  {Lecavelier des Etangs}, {H{\'e}brard}, {Ballester}, {Ehrenreich}, {Ferlet},
  {McConnell}, {Mayor}, \& {Parkinson}}]{2004ApJ...604L..69V}
{Vidal-Madjar}, A., {D{\'e}sert}, J.-M., {Lecavelier des Etangs}, A., {et~al.}
  2004, \apjl, 604, L69

\bibitem[{{Vidal-Madjar} {et~al.}(2013){Vidal-Madjar}, {Huitson}, {Bourrier},
  {D{\'e}sert}, {Ballester}, {Lecavelier des Etangs}, {Sing}, {Ehrenreich},
  {Ferlet}, {H{\'e}brard}, \& {McConnell}}]{2013A&A...560A..54V}
{Vidal-Madjar}, A., {Huitson}, C.~M., {Bourrier}, V., {et~al.} 2013, \aap, 560,
  A54

\bibitem[{{Vidal-Madjar} {et~al.}(2003){Vidal-Madjar}, {Lecavelier des Etangs},
  {D{\'e}sert}, {Ballester}, {Ferlet}, {H{\'e}brard}, \&
  {Mayor}}]{2003Natur.422..143V}
{Vidal-Madjar}, A., {Lecavelier des Etangs}, A., {D{\'e}sert}, J.-M., {et~al.}
  2003, \nat, 422, 143

\bibitem[{{Vidotto} \& {Bourrier}(2017)}]{2017MNRAS.470.4026V}
{Vidotto}, A.~A. \& {Bourrier}, V. 2017, \mnras, 470, 4026

\bibitem[{{Villarreal D'Angelo} {et~al.}(2021){Villarreal D'Angelo}, {Vidotto},
  {Esquivel}, {Hazra}, \& {Youngblood}}]{2021MNRAS.501.4383V}
{Villarreal D'Angelo}, C., {Vidotto}, A.~A., {Esquivel}, A., {Hazra}, G., \&
  {Youngblood}, A. 2021, \mnras, 501, 4383

\bibitem[{{Vissapragada} {et~al.}(2022){Vissapragada}, {Knutson},
  {Greklek-McKeon}, {Oklopcic}, {Dai}, {dos Santos}, {Jovanovic}, {Mawet},
  {Millar-Blanchaer}, {Paragas}, {Spake}, \& {Vasisht}}]{2022arXiv220411865V}
{Vissapragada}, S., {Knutson}, H.~A., {Greklek-McKeon}, M., {et~al.} 2022,
  arXiv e-prints, arXiv:2204.11865

\bibitem[{{Vissapragada} {et~al.}(2020){Vissapragada}, {Knutson}, {Jovanovic},
  {Harada}, {Oklop{\v{c}}i{\'c}}, {Eriksen}, {Mawet}, {Millar-Blanchaer},
  {Tinyanont}, \& {Vasisht}}]{2020AJ....159..278V}
{Vissapragada}, S., {Knutson}, H.~A., {Jovanovic}, N., {et~al.} 2020, \aj, 159,
  278

\bibitem[{{Vissapragada} {et~al.}(2021){Vissapragada}, {Stef{\'a}nsson},
  {Greklek-McKeon}, {Oklop{\v{c}}i{\'c}}, {Knutson}, {Ninan}, {Mahadevan},
  {Ca{\~n}as}, {Chachan}, {Cochran}, {Collins}, {Dai}, {David}, {Halverson},
  {Hawley}, {Hebb}, {Kanodia}, {Kowalski}, {Livingston}, {Maney}, {Metcalf},
  {Morley}, {Ramsey}, {Robertson}, {Roy}, {Spake}, {Schwab}, {Terrien},
  {Tinyanont}, {Vasisht}, \& {Wisniewski}}]{2021AJ....162..222V}
{Vissapragada}, S., {Stef{\'a}nsson}, G., {Greklek-McKeon}, M., {et~al.} 2021,
  \aj, 162, 222

\bibitem[{{Volkov} {et~al.}(2011){Volkov}, {Johnson}, {Tucker}, \&
  {Erwin}}]{2011ApJ...729L..24V}
{Volkov}, A.~N., {Johnson}, R.~E., {Tucker}, O.~J., \& {Erwin}, J.~T. 2011,
  \apjl, 729, L24

\bibitem[{{Waalkes} {et~al.}(2019){Waalkes}, {Berta-Thompson}, {Bourrier},
  {Newton}, {Ehrenreich}, {Kempton}, {Charbonneau}, {Irwin}, \&
  {Dittmann}}]{2019AJ....158...50W}
{Waalkes}, W.~C., {Berta-Thompson}, Z., {Bourrier}, V., {et~al.} 2019, \aj,
  158, 50

\bibitem[{{Wang} \& {Dai}(2018)}]{2018ApJ...860..175W}
{Wang}, L. \& {Dai}, F. 2018, \apj, 860, 175

\bibitem[{{Watson} {et~al.}(1981){Watson}, {Donahue}, \&
  {Walker}}]{1981Icar...48..150W}
{Watson}, A.~J., {Donahue}, T.~M., \& {Walker}, J.~C.~G. 1981, \icarus, 48, 150

\bibitem[{{Wyttenbach} {et~al.}(2015){Wyttenbach}, {Ehrenreich}, {Lovis},
  {Udry}, \& {Pepe}}]{2015A&A...577A..62W}
{Wyttenbach}, A., {Ehrenreich}, D., {Lovis}, C., {Udry}, S., \& {Pepe}, F.
  2015, \aap, 577, A62

\bibitem[{{Wyttenbach} {et~al.}(2020){Wyttenbach}, {Molli{\`e}re},
  {Ehrenreich}, {Cegla}, {Bourrier}, {Lovis}, {Pino}, {Allart}, {Seidel},
  {Hoeijmakers}, {Nielsen}, {Lavie}, {Pepe}, {Bonfils}, \&
  {Snellen}}]{2020A&A...638A..87W}
{Wyttenbach}, A., {Molli{\`e}re}, P., {Ehrenreich}, D., {et~al.} 2020, \aap,
  638, A87

\bibitem[{{Yan} {et~al.}(2021{\natexlab{a}}){Yan}, {Guo}, {Huang}, \&
  {Xing}}]{2021ApJ...907L..47Y}
{Yan}, D., {Guo}, J., {Huang}, C., \& {Xing}, L. 2021{\natexlab{a}}, \apjl,
  907, L47

\bibitem[{{Yan} \& {Henning}(2018)}]{2018NatAs...2..714Y}
{Yan}, F. \& {Henning}, T. 2018, Nature Astronomy, 2, 714

\bibitem[{{Yan} {et~al.}(2021{\natexlab{b}}){Yan}, {Wyttenbach},
  {Casasayas-Barris}, {Reiners}, {Pall{\'e}}, {Henning}, {Molli{\`e}re},
  {Czesla}, {Nortmann}, {Molaverdikhani}, {Chen}, {Snellen}, {Zechmeister},
  {Huang}, {Ribas}, {Quirrenbach}, {Caballero}, {Amado}, {Cont},
  {Khalafinejad}, {Khaimova}, {L{\'o}pez-Puertas}, {Montes}, {Nagel}, {Oshagh},
  {Pedraz}, \& {Stangret}}]{2021A&A...645A..22Y}
{Yan}, F., {Wyttenbach}, A., {Casasayas-Barris}, N., {et~al.}
  2021{\natexlab{b}}, \aap, 645, A22

\bibitem[{{Zhang} {et~al.}(2022{\natexlab{a}}){Zhang}, {Cauley}, {Knutson},
  {France}, {Kreidberg}, {Oklop{\v{c}}i{\'c}}, {Redfield}, \&
  {Shkolnik}}]{2022arXiv220402985Z}
{Zhang}, M., {Cauley}, P.~W., {Knutson}, H.~A., {et~al.} 2022{\natexlab{a}},
  arXiv e-prints, arXiv:2204.02985

\bibitem[{{Zhang} {et~al.}(2022{\natexlab{b}}){Zhang}, {Knutson}, {Dai},
  {Wang}, {Ricker}, {Schwarz}, {Mann}, \& {Collins}}]{2022arXiv220713099Z}
{Zhang}, M., {Knutson}, H.~A., {Dai}, F., {et~al.} 2022{\natexlab{b}}, arXiv
  e-prints, arXiv:2207.13099

\bibitem[{{Zhang} {et~al.}(2022{\natexlab{c}}){Zhang}, {Knutson}, {Wang},
  {Dai}, \& {Barrag{\'a}n}}]{2022AJ....163...67Z}
{Zhang}, M., {Knutson}, H.~A., {Wang}, L., {Dai}, F., \& {Barrag{\'a}n}, O.
  2022{\natexlab{c}}, \aj, 163, 67

\bibitem[{{Zhang} {et~al.}(2022{\natexlab{d}}){Zhang}, {Knutson}, {Wang},
  {Dai}, {dos Santos}, {Fossati}, {Henry}, {Ehrenreich}, {Alibert}, {Hoyer},
  {Wilson}, \& {Bonfanti}}]{2022AJ....163...68Z}
{Zhang}, M., {Knutson}, H.~A., {Wang}, L., {et~al.} 2022{\natexlab{d}}, \aj,
  163, 68

\bibitem[{{Zhang} {et~al.}(2021){Zhang}, {Knutson}, {Wang}, {Dai}, {Oklopcic},
  \& {Hu}}]{2021AJ....161..181Z}
{Zhang}, M., {Knutson}, H.~A., {Wang}, L., {et~al.} 2021, \aj, 161, 181

\bibitem[{{Zhang} {et~al.}(2020){Zhang}, {Snellen}, {Molli{\`e}re},
  {Alonso-Floriano}, {Webb}, {Brogi}, \& {Wyttenbach}}]{Zhang20}
{Zhang}, Y., {Snellen}, I.~A.~G., {Molli{\`e}re}, P., {et~al.} 2020, \aap, 641,
  A161

\end{thebibliography}

\end{document}